\documentclass[12pt,preprint]{aastex}

\newcommand{\beq}{\begin{equation}}
\newcommand{\eeq}{\end{equation}}
\newcommand{\beqa}{\begin{eqnarray}}
\newcommand{\eeqa}{\end{eqnarray}}
\newcommand{\bit}{\begin{itemize}}
\newcommand{\eit}{\end{itemize}}

\begin{document}

\title{Wouthuysen-Field Coupling in the 21 cm Region Around High Redshift Sources}

\author{Ishani Roy\altaffilmark{1}, Wen Xu\altaffilmark{2}, Jing-Mei
Qiu\altaffilmark{3},  Chi-Wang Shu\altaffilmark{1}, and Li-Zhi
Fang\altaffilmark{2}}

\altaffiltext{1}{Division of Applied Mathematics, Brown University,
Providence, RI 02912} \altaffiltext{2}{Department of Physics,
University of Arizona, Tucson, AZ 85721} \altaffiltext{3}{Department
of Mathematical and Computer Science, Colorado School of Mines,
Golden, CO 80401}

\begin{abstract}

The 21 cm emission and absorption from gaseous halos around the
first generation of stars substantially depend on the
Wouthuysen-Field (W-F) coupling, which relates the spin temperature
with the kinetic temperature of hydrogen gas via the resonant
scattering between Ly$\alpha$ photons and neutral hydrogen.
Therefore, the existence of Ly$\alpha$ photons in the 21 cm region
is essential. Although the center object generally is a strong
source of Ly$\alpha$ photons, the transfer of Ly$\alpha$ photons in
the 21 cm region is very inefficient, as the optical depth of Ly$\alpha$
photons is very large. Consequently, the Ly$\alpha$ photons $\nu_0$
from the source may not be able to transfer to the entire 21 cm
region timely to provide the W-F coupling. This problem is
especially important considering that the lifetime of first
stars generally is short. We investigate this problem with the
numerical solution of the integrodifferential equation, which
describes the kinetics of Ly$\alpha$ resonant photons in both
physical and frequency spaces. We show that the photon transfer
process in the physical space is actually coupled to that in the
frequency space. First, the diffusion in the frequency space
provides a shortcut for the diffusion in the physical space. It
makes the mean time for the escape of resonant photon in optical
depth $\tau$ media roughly proportional to the optical depth $\tau$,
not $\tau^2$. Second and more importantly, the resonant scattering
is effective in bouncing photons with frequency $\nu\neq \nu_0$ back
to $\nu_0$. This process can quickly restore $\nu_0$ photons and
establish the local Boltzmann distribution of the photon spectrum
around $\nu_0$. Therefore, the mechanism of ``escape via shortcut''
plus ``bounce back'' enables the W-F coupling to be properly
realized in the 21 cm region around first stars. This mechanism
also works for photons injected into the 21 cm region by redshift.

\end{abstract}


\newpage

\section{Introduction}

The resonant scattering of Ly$\alpha$ photons with neutral hydrogen atoms leads to
a local Boltzmann distribution of the photon spectrum around the Ly$\alpha$ frequency
$\nu_0$ with color temperature equal to the kinetic temperature $T$ of hydrogen gas.
Consequently, the spin temperature $T_s$ of the hyperfine structure of neutral hydrogen will
be coupled to the kinetic temperature of hydrogen gas. This is the so-called
Wouthuysen-Field (W-F) coupling (Wouthuysen 1952; Field 1958, 1959). The W-F coupling
is crucial to estimate the redshifted 21 cm signal from the halos of first generation
of stars, because the deviation of the spin temperature from the temperature of cosmic
microwave background (CMB) $T_{\rm cmb}$ is considered to be mainly caused by the
W-F coupling (e.g. Furlanetto et al. 2006).

The lifetime of the first stars is short. The ionized and heated
halos around the first luminous objects are strongly time-dependent.
The time scale of the evolution of the expected 21 cm
emission/absorption regions can be as small as 10$^5$ years  (Cen
2006; Liu et al. 2007). Therefore, the 21 cm signal models based on
the W-F coupling would be reasonable only if the time scale of the
onset of the W-F coupling is less than that of the 21 cm region
evolution. This time evolution has been studied very recently
(Roy et al. 2009a, hereafter referred to as PaperI). It concludes that the local
Boltzmann distribution can form within a time scale shorter than 10$^5$
years, but it would take 10$^{5}$ years or even longer to reach its saturation
(time-independent) state.  Therefore, it is legitimate
to assume that the W-F coupling is taking place in the 21 cm
emission/absorption regions, but the intensity of the photon flux is less
than the time-independent solution.

PaperI studied, however, only the case of spatial homogeneity and
isotropy. It is equivalent to assume that the Ly$\alpha$
photons are uniformly distributed in the entire 21 cm signal region. This
assumption is not trivial. All the Ly$\alpha$ photons in the 21 cm
regions come from first stars, either from direct emission of Ly$\alpha$
photons, or from the Hubble redshifted photons. On the other hand, the 21
cm regions are highly opaque for Ly$\alpha$ photons. The W-F
coupling may not be uniformly available in the 21 cm signal region,
if the time scale of the transfer of Ly$\alpha$ photons in the physical
space is longer than that of the evolution of the 21 cm region.

The problem of Ly$\alpha$ photon transfer in optical thick media is
not new. It has been addressed as early as 1960s in relation to the
escape of resonant photons from opaque clouds (Osterbrock 1962;
Harrington 1973; Avery \& House 1968; Adams 1972). In these references
it is shown  that the diffusion of the photon distribution in the frequency
space caused by the resonant scattering will be helpful to speed up the spatial
diffusion. In the past decade, there are also many works on the
escape of Ly$\alpha$ photons from high redshift objects with the
mechanism of Hubble redshift (e.g. Miralda-Escude \& Rees 1998; Loeb \&
Rybicki 1999; Zheng \& Miralda-Escude 2002; Haiman \& Cen 2005; Tasitsiomi
2006; McQuinn et al. 2007). The Hubble redshifted Ly$\alpha$ photons are
also easy to take spatial transfer.  Although both the resonant scattering
and the Hubble redshift are useful to solve the problem of
Ly$\alpha$ photon transfer in the 21 cm region, they rely on the change of
photon frequency from $\nu_0$ to $\nu_0\pm \Delta \nu$. Therefore, the
local Boltzmann frequency distribution will be disturbed, even if it
initially is in the state of local Boltzmann frequency distribution.
Thus, it is unclear whether the W-F coupling keeps to work timely
and uniformly in the 21 cm region.

In this context, a time-dependent solution of the kinetics of
Ly$\alpha$ photons in both physical and frequency spaces of the 21
cm region is necessary. This is the topic of the current paper. We will
show that the resonant scattering of Ly$\alpha$ photons is effective
to solve the problem of spatial transfer in optical thick 21 cm
region as well as to keep the W-F coupling working. On the other
hand, for 21 cm regions of short lifetime objects, the cosmic
expansion does not provide effective mechanism for the spatial
transfer of Ly$\alpha$ photons.

Similar to PaperI, we use the numerical solution of the
time-dependent integrodifferential equation of the radiative transfer with resonant
scattering. The numerical solver is based on the weighted essentially non-oscillatory
(WENO) scheme (Jiang \& Shu 1996). WENO scheme is effective in solving Boltzmann
equations (Carrillo et al. 2003, 2006) and radiative transfer (Qiu et al. 2006, 2007,
2008). The algorithm related to resonant scattering has also been given in
Roy et al. (2009b). This numerical solver has successfully passed the tests of
analytic solutions and conservation of photon number. Therefore, it is a good
candidate for computing the current problem.

This paper is organized as follows. Section 2 addresses the physical
problems of Ly$\alpha$ photon transfer in the 21 cm emission and
absorption regions of luminous objects. Section 3 is on the time
scale of the transfer of resonant scattering in the frequency space.
Section 4 presents Ly$\alpha$ photon transfer in the frequency and
physical spaces. These results can be used for the 21 cm signal
region (\S 5). Discussion and conclusion are given in Section 6. The
details of the numerical implementation are given in the Appendix.

\section{Radiative transfer problem of the 21 cm region}

\subsection{Basic properties of the 21 cm region}

The property of the ionized and heated regions around an individual
luminous object is dependent on the luminosity (or mass), the
spectrum of UV photon emission, and the time-evolution of the center
object (Cen 2006; Liu et al. 2007). We will not work on a specific
model, but consider only the common features. These halos
generally consist of three spheres. The most inner region of the
halo is the highly ionized Str\"omgren sphere, or the HII region, in
which the fraction of neutral hydrogen $f_{\rm HI}=n_{\rm HI}/n_{\rm
H}$ is no more than 10$^{-5}$, where $n_{\rm HI}$ and $n_{\rm H}$
are, respectively, the number densities of neutral hydrogen HI and
total hydrogen. The temperature of the HII region is about 10$^4$ K.
The physical radius of this sphere is in the range of a few to
a few tens of kpc. The second region is the 21 cm emission shell, which is
just outside the HII region. The physical size of this shell is similar
to that of the HII sphere. The temperature of hydrogen gas in
the emission shell is in the range $10^2 < T< 10^4$ K due to the
heating of UV photons. The fraction $f_{\rm HI}$ is in the range of 0.1
to 1. The third region is the 21 cm absorption shell, which is
outside the 21 cm emission region. The physical size is on the order of 100 kpc.
The temperature of this region is lower than $T_{\rm cmb}=T_{\rm CMB}(1+z)$,
where $T_{\rm CMB}$ is the temperature of CMB today. The time scale of the
formation of the halos is about 10$^6$ years. The lifetime of the halos is
about the same as the lifetime of the first stars.

At the epoch of redshift $z\simeq 20$, the reionization region consists of isolated
patches around first sources. Most Ly$\alpha$ photons in the 21 cm regions
should come from the central source and the subsequent re-emission processes. If
one can estimate the center object as a Ly$\alpha$ emitter, the emission of Ly$\alpha$
photons
in number per unit time would be about $dN_{\rm Ly\alpha}/dt = 10^{53}$ s$^{-1}$.
The recombination of HII and electron in the Str\"omgren sphere is also a source
of Ly$\alpha$ photons. If the physical radius of the Str\"omgren sphere is
$\sim 10$ kpc, the emission intensity is about $dN_{\rm Ly\alpha}/dt = 10^{49}$
s$^{-1}$.

Using the parameters of the concordance $\Lambda$CDM model, the optical depths of
photons with Ly$\alpha$ resonant frequency $\nu_0$ in the 21 cm region are
\begin{equation}
\label{eq1}
\tau(\nu_0)=n_{\rm HI}\sigma_0R=4.9 \times 10^6 f_{\rm HI}\left (
\frac{T}{10^4}\right
)^{-1/2}\left(\frac{1+z}{20}\right)^3\left(\frac{\Omega_bh^2}{0.022}\right)
\left (\frac{R}{{\rm 10 kpc}}\right),
\end{equation}
where $\sigma_0$ is the cross section of the resonant scattering at
the frequency $\nu_0$,  $R$ is the physical size of the considered
sphere, and $\tau$ is the distance $R$ in the units of
the mean free path of $\nu_0$ photons at redshift $z$. Eq.(\ref{eq1}) shows that
the optical depth of the 21 cm signal regions with $f_{\rm HI}\geq 0.1$ is
$\tau(\nu_0)\geq 10^6$.

It is also useful to define a dimensionless time
$\eta=cn_{\rm HI}\sigma_0 t$ as
\begin{equation}
\label{eq2}
t=\eta/cn_{\rm HI}\sigma_0=6.7\times 10^{-3}f^{-1}_{\rm
HI}\left(\frac{T}{10^4}\right)^{1/2}\left (\frac{20}{1+z}\right
)^3\left (\frac{0.022}{\Omega_b h^2}\right )\eta \hspace{3mm} yrs.
\end{equation}
This is the time in the units of mean free flight-time. For a time scale
$t\simeq 10^5$ yrs, the scale of $\eta$ is $\simeq 10^7$.

\subsection{Problems with the W-F coupling}

There are, at least, two radiative transfer problems with the W-F coupling
in the 21 cm regions: A.) How to provide enough Ly$\alpha$ photons in such opaque
medium? B.) Can the radiative transfer in the 21 cm region keep the
W-F coupling to work well?

If photons {\it always} keep the frequency $\nu_0$, their spatial
diffusion can be described as a random walk process (Chandrasekar
1943). The mean number of scattering required for the diffusion over a
range $R$ is on the order of $\tau^2$. Thus, the time for the diffusion over the 21 cm
range is $\eta\simeq \tau^2$, corresponding to the time
$t = \tau^2 /cn_{\rm HI}\sigma_0 \geq  10^{12}$ years. This diffusion
mechanism, obviously, is useless for the 21 cm regions.

The time scale of the diffusion would be substantially reduced if the
frequency of the Ly$\alpha$ photons can have a small shift from $\nu_0$ to
$\nu = \nu_0\pm \Delta \nu$, because the optical depth $\tau$ at the frequency
$\nu_0\pm \Delta \nu$ is significantly less than $\tau(\nu_0)$. One can estimate
the frequency shift $\Delta \nu$ by the condition of optical depth
$\tau(\nu_0\pm \Delta \nu)\simeq 1$, which is
\begin{equation}
\label{eq3}
\tau(\nu_0)\frac{\phi(x,a)}{\phi(0,a)} \leq 1
\end{equation}
where $\phi(x,a)$ is the Voigt function for the resonant line $\nu_0$ profile as
(Hummer 1965)
\begin{equation}
\label{eq4}
\phi(x,a)=\frac{a}{\pi^{3/2}}\int^{\infty}_{-\infty} dy \frac
{e^{-y^2}}{(x-y)^2+a^2}
\end{equation}
where the dimensionless variable $x$ is defined by $x=(\nu-\nu_0)/\Delta
\nu_D$ and $\Delta \nu_D=\nu_0v_T/c$ is the Doppler broadening of
hydrogen gas with thermal velocity $v_T=\sqrt{k_bT/2m_{\rm H}}$. Therefore, $x$
measures the frequency deviation
$\Delta \nu=|\nu-\nu_0|$ in the units of the Doppler broadening. The
parameter $a$ in eq.(\ref{eq4}) is the ratio of the natural to the Doppler
broadening. For Ly$\alpha$ line, $a=2.35\times
10^{-4}(T/10^4)^{-1/2}$. In terms of $x$, the solution of eq.(\ref{eq3})
is $x \geq 3$.

An effective mechanism of the frequency shift is given by the diffusion of the photon
distribution in the frequency space. Considering this mechanism, the number of
scattering required for diffusion over a physical distance $R$ is no longer equal to
$\tau^2$, but roughly on the order of $\tau$ (Osterbrock 1962; Harrington 1973;
Avery \& House 1968; Adams 1972). Since $\tau \propto R$, the time scale of
the spatial diffusion over size $R$ is comparable to $R/c$. Problem A would then be
solved.

Problem B still remains. If photons with the frequency $\nu <
\nu_0 - \Delta \nu$ or $\nu > \nu_0 + \Delta \nu$ take a faster
spatial diffusion than the $\nu_0$ photons, how
can we keep the W-F coupling to work? Without $\nu_0$ photons, one cannot have
a local Boltzmann distribution around $\nu_0$. Therefore, the
photons with the frequency $\nu < \nu_0 - \Delta \nu$ or $\nu > \nu_0 +
\Delta \nu$ should be brought back to the frequency $\nu_0$.
We must study whether the frequency space diffusion mechanism can
restore $\nu_0$ photons from  photons with the frequency $\nu\neq \nu_0$.

Cosmic expansion also leads to a deviation of photon frequency from
$\nu_0$ to a lower one $\nu_0-\Delta\nu$, which speeds up the spatial transfer.
However, we also need  a mechanism to restore Ly$\alpha$  photons from
Hubble redshifted photons. Therefore, in terms of the W-F coupling in the 21 cm region,
we must study the kinetics of Ly$\alpha$ photons in both physical and
frequency spaces.

\section{Radiative transfer of resonant photons in the frequency space}

\subsection{Equations}

We first estimate the time scale needed for the frequency shift. We can use
the radiative transfer equation of a homogeneous and isotropically expanding infinite
medium consisting of neutral hydrogen. The equation of the mean intensity $J$ in
terms of the photon number is (Hummer \& Rybicki 1992; Rybick \& Dell'antonio 1994)
\begin{eqnarray}
\label{eq5}
\frac{\partial J(x, \eta)}{\partial  \eta}& =& - \phi(x,a)J(x,  \eta) \nonumber \\
 & & +\int \mathcal{R}(x,x';a)J(x', \eta)dx' + \gamma \frac{\partial J}{\partial x}
+S(x, \eta).
\end{eqnarray}
 The parameter
$\gamma^{-1}$ measures the number of scattering during a Hubble
time, given by
\begin{equation}
\label{eq6}
\gamma^{-1}= 1.4 \times 10^6 h^{-1}f_{\rm
HI}\left(\frac{0.25}{\Omega_M}\right
)^{1/2}\left(\frac{\Omega_bh^{2}}{0.022}\right )
\left(\frac{1+z}{20}\right )^{3/2}.
\end{equation}
The re-distribution function $\mathcal{R}(x,x')$ gives the
probability of a photon absorbed at the frequency $x'$, and
re-emitted at the frequency $x$. It depends on the details of the
scattering (Henyey 1941; Hummer 1962; Hummer, 1969). If we consider
coherent scattering without recoil, the re-distribution function with the
Voigt profile eq.(\ref{eq4}) is
\begin{eqnarray}
\label{eq7}
\lefteqn{ \mathcal{R}(x,x';a)= } \\ \nonumber
 & \ \ \  & \frac{1}{\pi^{3/2}}\int^{\infty}_{|x-x'|/2}e^{-u^2}
\left [
\tan^{-1}\left(\frac{x_{\min}+u}{a}\right)-\tan^{-1}\left(\frac{x_{\max}-u}{a}\right
)\right ]du
\end{eqnarray}
where $x_{\min}=\min(x, x')$ and $x_{\max}=\max(x,x')$. In the case of $a=0$, i.e.
considering only the Doppler broadening, the re-distribution function is
\begin{equation}
\label{eq8}
\mathcal{R}(x,x')=\frac{1}{2}e^{2bx'+b^2}{\rm
erfc}[{\rm max}(|x+b|,|x'+b|)],
\end{equation}
where the parameter $b=h\nu_0/mv_T c= 2.5\times 10^{-4} (10^4/T)^{1/2}$
is due to the recoil of atoms. The re-distribution function of eq.(\ref{eq8}) is normalized as
$\int_{-\infty}^{\infty} \mathcal{R}(x,x')dx'=\phi(x,0)\equiv\phi_g(x)
=\frac{1}{\sqrt{\pi}}e^{-x^2}$.
The numerical algorithm to solve eq.(\ref{eq5}) has been given in Roy et al. (2009a, 2009b).

\subsection{Time scales of the frequency shift}

\begin{figure}[htb]
\centering
\includegraphics[width=8cm]{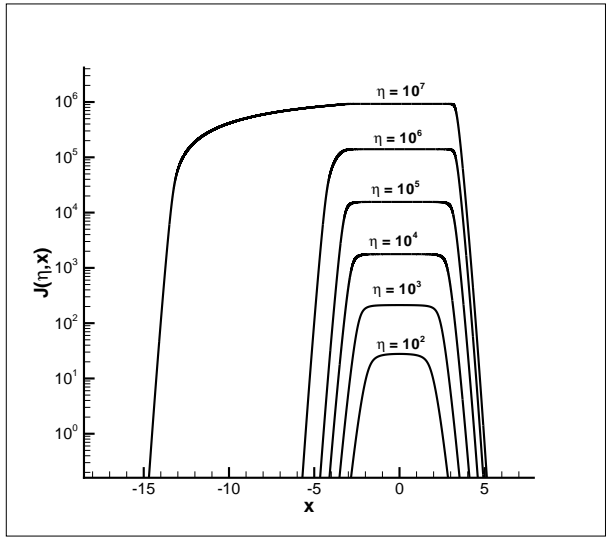}
\caption{Solutions $J(\eta,x)$ of eq. (\ref{eq5}) with the re-distribution
function eq.(\ref{eq8}).
Parameters $b=0$, and $\gamma=10^{-6}$.}
\label{fig1}
\end{figure}

On the right hand side of eq. (\ref{eq5}), the first term is the absorption at
the resonant frequency $x$, the second term is the re-emission of photons with
frequency $x$ by scattering, and the third term describes the Hubble redshift of photons.
We first solve eq. (\ref{eq5}) by dropping the terms of
absorption and re-emission. The equation is
\begin{equation}
\label{eq9}
\frac{\partial J}{\partial \eta}= \gamma\frac{\partial J}{\partial
x}+S(x,\eta).
\end{equation}
Assuming the source is $S=C\phi_s(x)$, where $\phi_s(x)$ is the normalized frequency
profile of the source photons. The analytic solution of eq.(\ref{eq9}) is
(Rybicki \& Dell'antonio, 1994)
\begin{equation}
\label{eq10}
J(x,\eta)=J(x+\gamma\eta,0) +
C\gamma^{-1}\int^{x+\gamma
\eta}_{x}\phi_s(x') dx'
\end{equation}
where $J(x, 0)$ is the initial flux. Define the mean frequency by
\begin{equation}
\label{eq11}
\bar{x}(\eta)\equiv \frac{\int x J(\eta, x)dx}{\int J(\eta, x)dx}.
\end{equation}
If the initial flux is $J(x, 0)=0$, one can show that for any
profile $\phi_s(x)$ we have
\begin{equation}
\label{eq12}
\bar{x}(\eta)=- \gamma \eta /2.
\end{equation}
As expected, the speed of redshift $d\bar{x}/d\eta$ is a constant.
For the Hubble expansion, a frequency shift $x$ needs  a time
$\eta \simeq \gamma^{-1}x$. Thus, in order to have the frequency
shift $x \geq 3$, the time scale is $\eta\geq 6\times 10^6$,
corresponding to $t\geq 4\times 10^4$ years. This scale seems to be short enough
compared with the lifetime of first stars. However, Hubble redshift is
less effective comparing with the resonant scattering. This point can be
seen in Figure \ref{fig1}, which gives the solution of eq.(\ref{eq5}) with
the re-distribution function eq.(\ref{eq8}). The source is taken to be
$S(x)= \phi_g(x) =(1/\sqrt{\pi})e^{-x^2}$. We also take the parameter $\gamma$
to be $10^{-6}$ [eq.(\ref{eq6})], and ignore recoil, i.e. $b=0$.

Figure \ref{fig1} shows that the diffusion in the frequency space leads to
a flat plateau with width $|x|\leq 3$ when $\eta$ is as small as
$\eta \simeq 10^4$, or $t\simeq 10^2$ years. This time scale is much less
than that of the Hubble redshift. Therefore, the major
mechanism to produce photons with frequency $|x| \geq 3$  is given by the
resonant scattering.

\subsection{Bounce back mechanism}

 From Figure \ref{fig1}, we can see that the profile of photons are almost
symmetric with respect to $x=0$ (or to $\nu=\nu_0$) until about the time
$\eta=10^6$. The redshift effect can only be seen from the curve of $\eta \geq 10^6$.
That is, the resonant scattering impedes cosmic redshift. The
impediment is due to the ``bounce back'' of resonant scattering.
Regardless of whether the frequency of the absorbed photons is larger or smaller
than $\nu_0$, the mean frequency of the re-emitted photons is always $\nu_0$.
Therefore, the resonant scattering will bring back some redshifted photons to
the frequency $\nu_0$. Thus the net effect of resonant scattering
(absorption and re-emission) is, on average, to bounce redshifted
photons back to the resonant frequency $\nu_0$, and to restore the
symmetry with respect to $x=0$. This bounce back mechanism is the key of restoring
$\nu_0$ photons and the W-F coupling
(see \S 4).

\begin{figure}[htb]
\centering
\includegraphics[width=8cm]{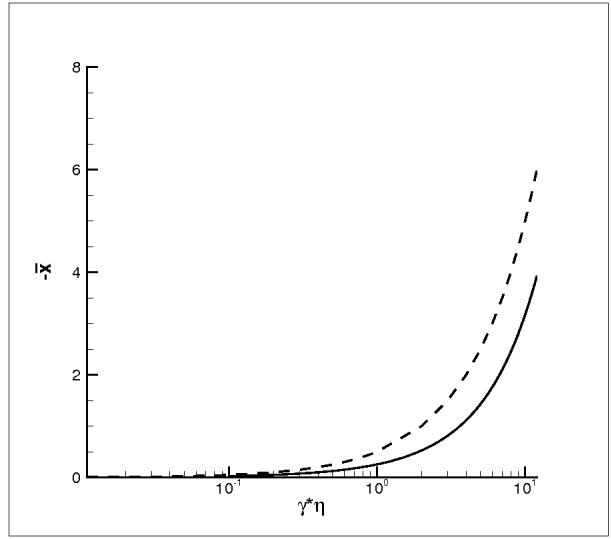}
\caption{$\bar{x}$ vs. $\gamma \eta$ for the solution shown in Figure
\ref{fig1} (solid) and the solution of eq.(\ref{eq12}) (dashed).}
\label{fig2}
\end{figure}

To illuminate the bounce back effect, we calculate the mean
frequency $\bar{x}(\eta)$ for the solutions shown in Figure \ref{fig1}. The
result is plotted in Figure \ref{fig2}. The solution of pure cosmic redshift
eq.(\ref{eq12}) is also shown in Figure \ref{fig2} as the dashed curve.
We can see from Figure \ref{fig2} that the
speed of photon frequency shift $d\bar{x}/d\eta$ when considering
resonant scattering is in general less than that in the case of pure cosmic
redshift. When the time $\eta$ is large, many photons
have been redshifted out of the frequency range of $\phi_g(x)$. In this case, the
``bounce back'' effect ceases, and redshift speed is recovered to
$d\bar{x}/d\eta=-\gamma/2$. From Figure \ref{fig2} one can see once again that
the Hubble redshift can produce frequency shift from $x=0$ to $|x|\simeq 3$ only
when $\eta \geq 10^7$, or $t\simeq 10^5$ years. This time scale may not be short
enough to match the 21 cm region with a short lifetime. When the time scale of
resonant scattering is less than the time scale of Hubble expansion,
the frequency shift of the Hubble expansion slows down significantly
by the resonant scattering.

\section{Radiative transfer (RT) of Ly$\alpha$ photons in the 21 cm regions}

\subsection{RT equation in spherical halo}

Considering a photon source located at the central region of a
uniformly distributed expanding medium, we can use the RT equation
of the specific intensity $I(\eta,r,x,\mu)$ as follows
\begin{eqnarray}
\label{eq13}
\lefteqn{ {\partial I\over\partial \eta} + \mu \frac{\partial I}
{\partial r}+\frac{(1-\mu^2)}{r}\frac{\partial I}{\partial \mu}
- \gamma \frac{\partial I}{\partial x} = } \nonumber \\
 & &   - \phi(x)I + \int \mathcal{R}(x,x')I(\eta, r,
x',\mu')dx'd\mu' + S,
\end{eqnarray}
where $\mu=\cos \theta$ is the direction relative to the radius
vector ${\bf r}$. The dimensionless coordinate $r$ is rescaled from
the physical coordinate $r_p$ as
\begin{equation}
\label{eq14}
r_p=2.1\times 10^{-3} f^{-1}_{\rm HI} \left
(\frac{T}{10^4}\right )^{1/2} \left (\frac{20}{1+z}\right )^3 \left
(\frac{0.022}{\Omega_b h^2}\right ) r \hspace{3mm} {\rm pc} .
\end{equation}
With these variables, the propagation of a signal with the speed of light will be
described by the equation $r= \eta+ {\rm const}$. It would still be reasonable
to use the isotropic approximation of the re-distribution (Mihalas et al. 1976).

When the optical depth is large, the Eddington approximation would
be proper. It is
\begin{equation}
\label{eq15}
I(\eta,r,x,\mu)\simeq J(\eta, r, x) + 3\mu F(\eta,r,x)
\end{equation}
where $J(\eta,r,x)=\frac{1}{2}\int_{-1}^{+1}I(\eta,r,x,\mu)d\mu$ is
the angularly averaged specific intensity and
$F(\eta,r,x)=\frac{1}{2}\int_{-1}^{+1}\mu I(\eta,r,x,\mu)d\mu$ is the flux.
Defining $j=r^2J$ and $f=r^2F$, Eq.(\ref{eq13})  yields the equations of $j$
and $f$ as
\begin{eqnarray}
\label{eq16}
{\partial j\over\partial \eta} + \frac{\partial f} {\partial r} & =
& - \phi(x)j + \int
\mathcal{R}(x,x')j dx'+  \gamma \frac{\partial
j}{\partial x}+ r^2S,\\
\label{eq17}
\frac{\partial f}{\partial \eta} + \frac{1}{3} \frac{\partial j}
{\partial r} - \frac{2}{3}\frac{j}{r} & = & - \phi(x)f.
\end{eqnarray}
The numerical algorithm for eqs.(\ref{eq16}) and (\ref{eq17}) is given in the Appendix.

Since photons with frequency shifted away from $\nu_0$ would be optical
thin, the equations (\ref{eq16}) and (\ref{eq17}) will no longer be
a good approximation when the frequency of photons is shifted to the optical
thin case. From Figures \ref{fig1} and \ref{fig2}, one can see that within
$\eta \leq 10^6$, most photons are still trapped in the frequency $|x|\leq 3$,
for which the optical depth is larger than 1. Therefore, the
Eddington approximation would be proper, at least, until $\eta$ is as large
as about $10^6$.

\subsection{Diffusion in the physical space}

\begin{figure}[htb]
\centering
\includegraphics[width=7cm]{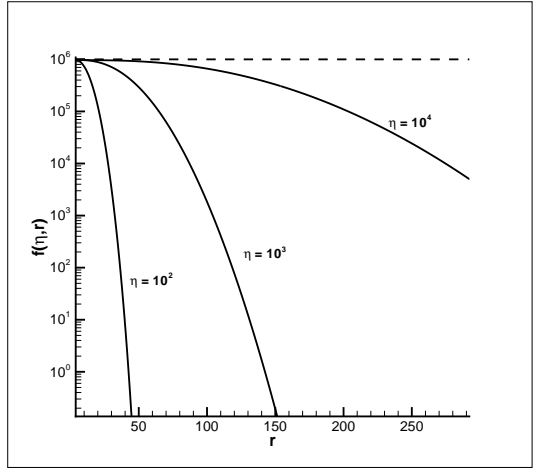}
\caption{The solution $f(\eta, r)$ of eqs.(\ref{eq18}) and
(\ref{eq19}). The straightline is $f=S_0=10^6$. }
\label{fig3}
\end{figure}

We first solve the equations (\ref{eq16}) and (\ref{eq17}) by
dropping all terms on the transfer in frequency.  It has been shown
that the source term can be replaced by a boundary condition of $f$
and $j$ at $r=r_0$ (Qiu et al. 2006). The equations then become
\begin{eqnarray}
\label{eq18}
{\partial j\over\partial \eta} + \frac{\partial f} {\partial r} & =
& 0,\\
\label{eq19}
{\partial f\over\partial \eta} + \frac{1}{3} \frac{\partial j}
{\partial r} - \frac{2}{3}\frac{j}{r}& = & - f.
\end{eqnarray}
We take the boundary condition at $r_0=1$ to be
\begin{equation}
\label{eq20}
j(\eta, 1)=3S_0, \hspace{1cm} f(\eta, 1)=S_0=10^6.
\end{equation}
Since eqs.(\ref{eq18}) and (\ref{eq19}) are linear, the parameter
$S_0$ is not important if we are only interested in the shape of
 $j$ and $f$ as function of $\eta$ and $r$. The initial condition is
taken to be
\begin{equation}
\label{eq21}
j(0,r)=f(0,r)=0.
\end{equation}
The solution of $f(\eta, r)$ is presented in Figure \ref{fig3}. The
dashed line is $f=S_0=10^6$, which is the time-independently exact
solution of eq.(\ref{eq18}). It is a horizontal straight line
because the flux $f=r^2F$ is $r$-independent and satisfies the
conservation of the photon number. Figure \ref{fig3} shows that the
spatial size $r$ of $f(\eta, r)$ roughly satisfies $r\propto
\sqrt{\eta}$. Therefore, without resonant scattering, the diffusion
basically is a random walk process as discussed in \S 2.2.

\begin{figure}[htb]
\centering
\includegraphics[width=5cm]{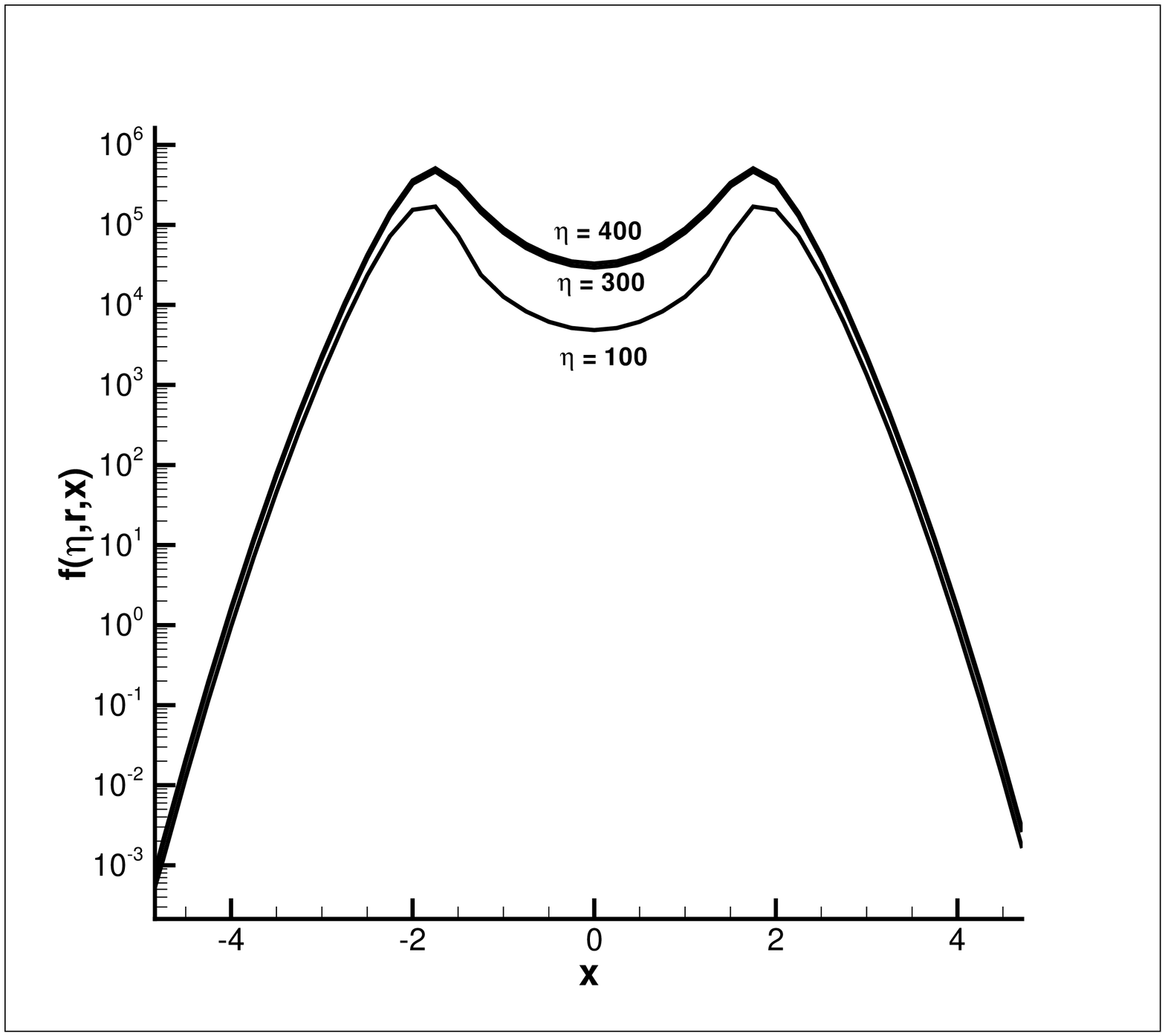}
\includegraphics[width=5cm]{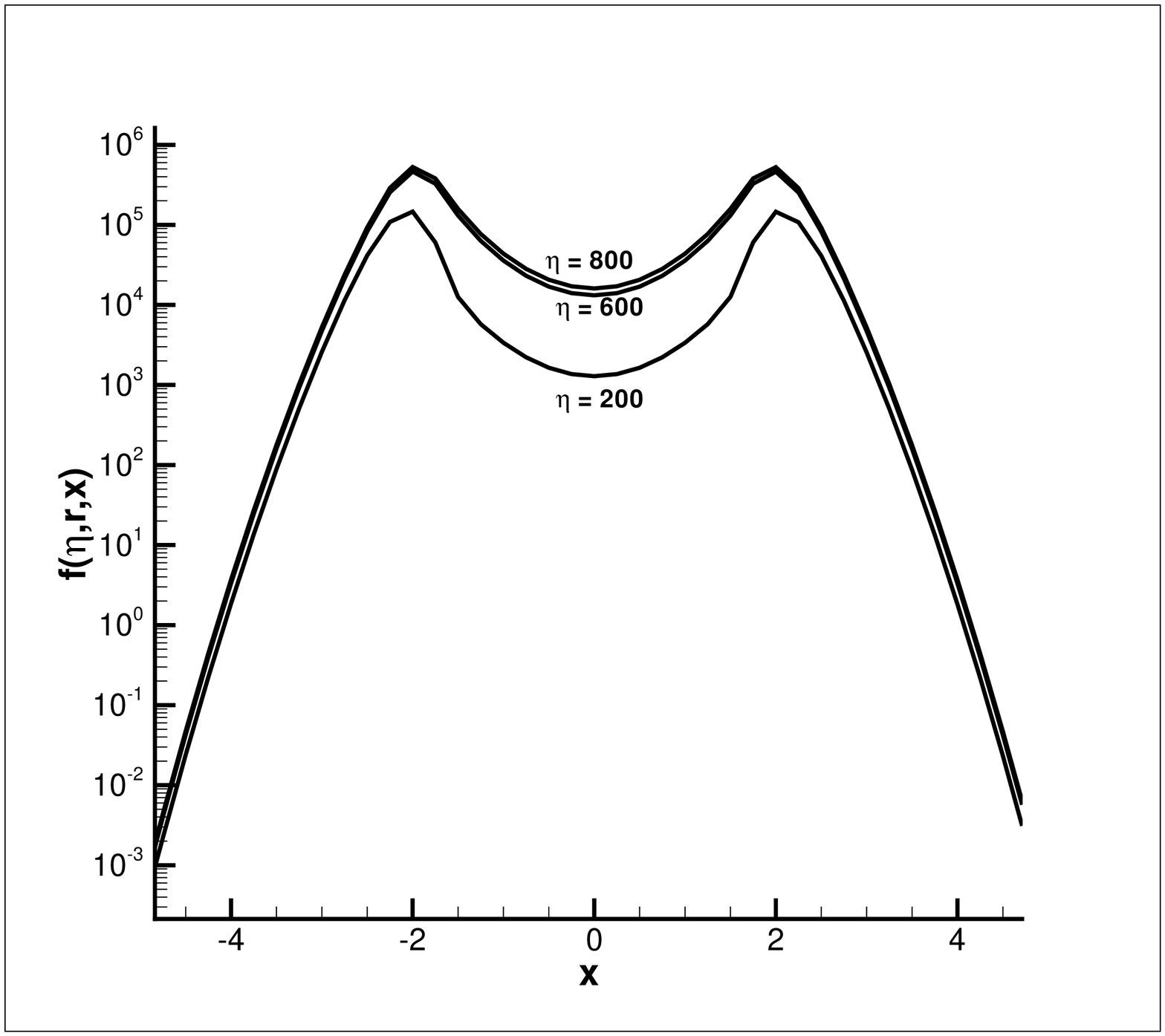}
\includegraphics[width=5cm]{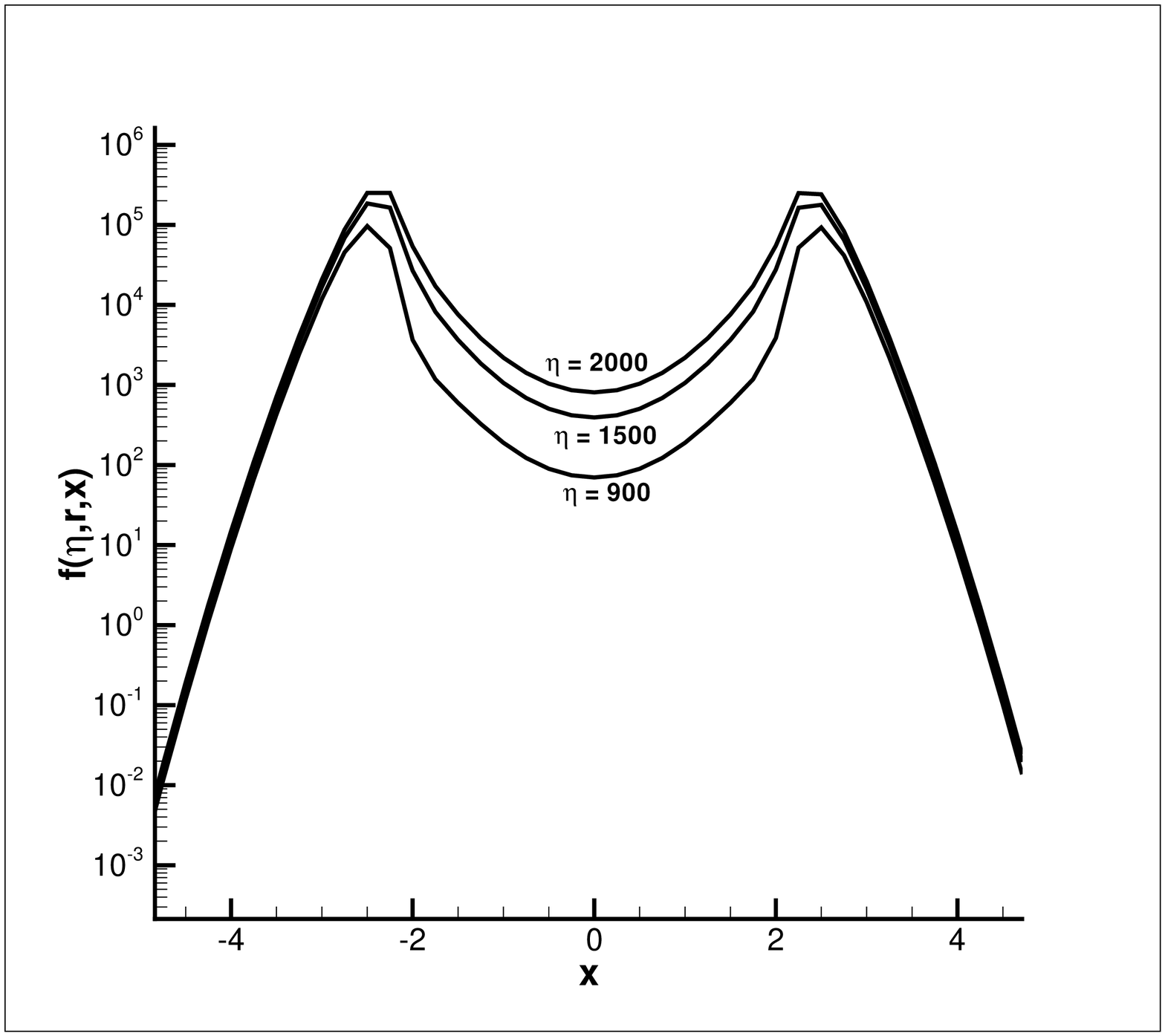}
\caption{The flux $f(\eta,r,x)$ of the solutions of eqs.(\ref{eq16}) and (\ref{eq17})
at $r=$ 50 (left), 100 (middle) and 500 (right).  The frequency profile of the
source is $\phi_g(x)=(1/\sqrt{\pi})e^{-x^2}$.
}
\label{fig4}
\end{figure}

We now turn to the solution of eqs.(\ref{eq16}) and (\ref{eq17})
with resonant scattering and Hubble redshift. The relevant
parameters are given by $b=0$ and $\gamma=10^{-5}$. We use the
boundary condition at $r=0$
\begin{equation}
\label{eq22}
j(\eta, 0, x)=0, \hspace{1cm} f(\eta, 0, x)=S_0 \phi_s(x)
\end{equation}
where the frequency profile is taken to be the Gaussian profile, i.e.
$\phi_s(x)=\phi_g(x)$. The parameter $S_0$ is still taken to be
$10^6$. The initial condition is similar to eq.(\ref{eq21}), i.e.
$j(0,r,x)=f(0,r,x)=0$. The solutions of $f(\eta,r,x)$ are plotted in Figure
\ref{fig4}.

All the solutions of Figure \ref{fig4} show two remarkable peaks at $x\simeq
\pm (2-3)$ for all radius $r$. The amplitude of $f(\eta,r,x)$ at the
peaks is higher than that at $x=0$ by a factor of 10 to 10$^2$. It
shows that the flux is dominated by photons with frequency $x\simeq
\pm(2-3)$. That is, the spatial transfer is carried out  by photons of $x\simeq
\pm (2-3)$. The amplitude of the flux at the saturated peaks is
basically $r$-independent, and the saturated values of $\int f(\eta,
r, x)dx$ are $r$-independent. This is consistent with the
conservation of the photon number.

\begin{figure}[htb]
\centering
\includegraphics[width=5cm]{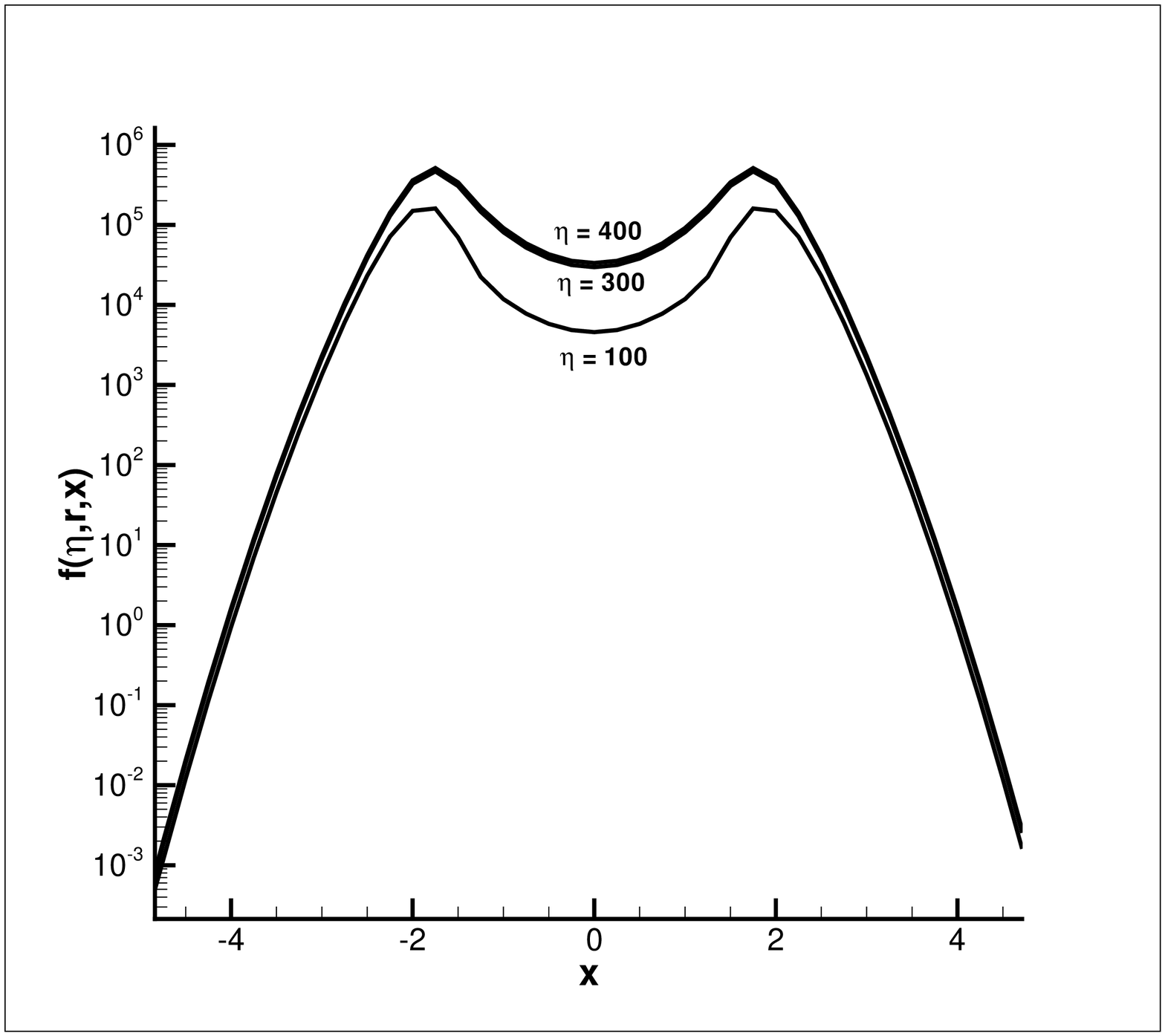}
\includegraphics[width=5cm]{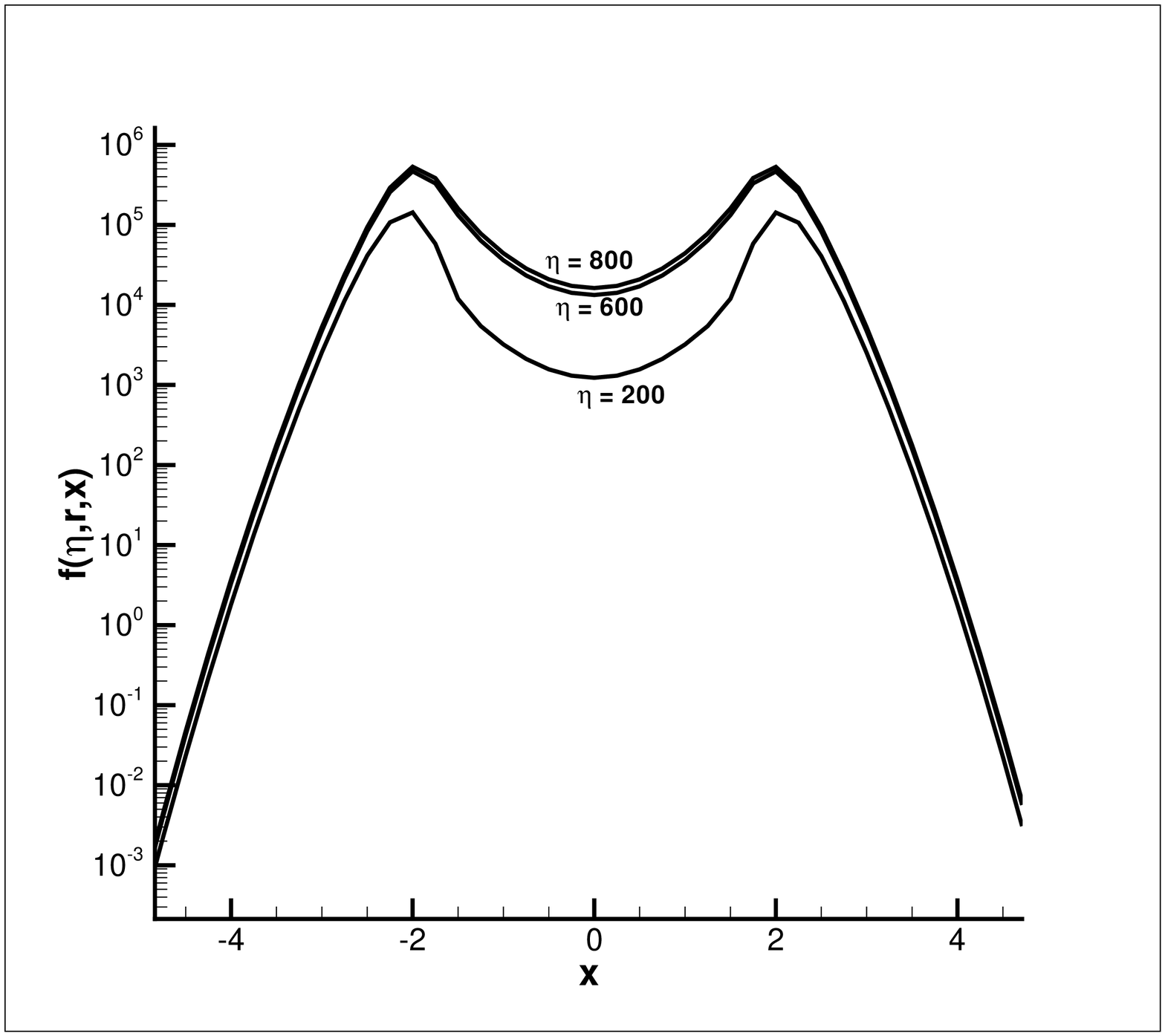}
\includegraphics[width=5cm]{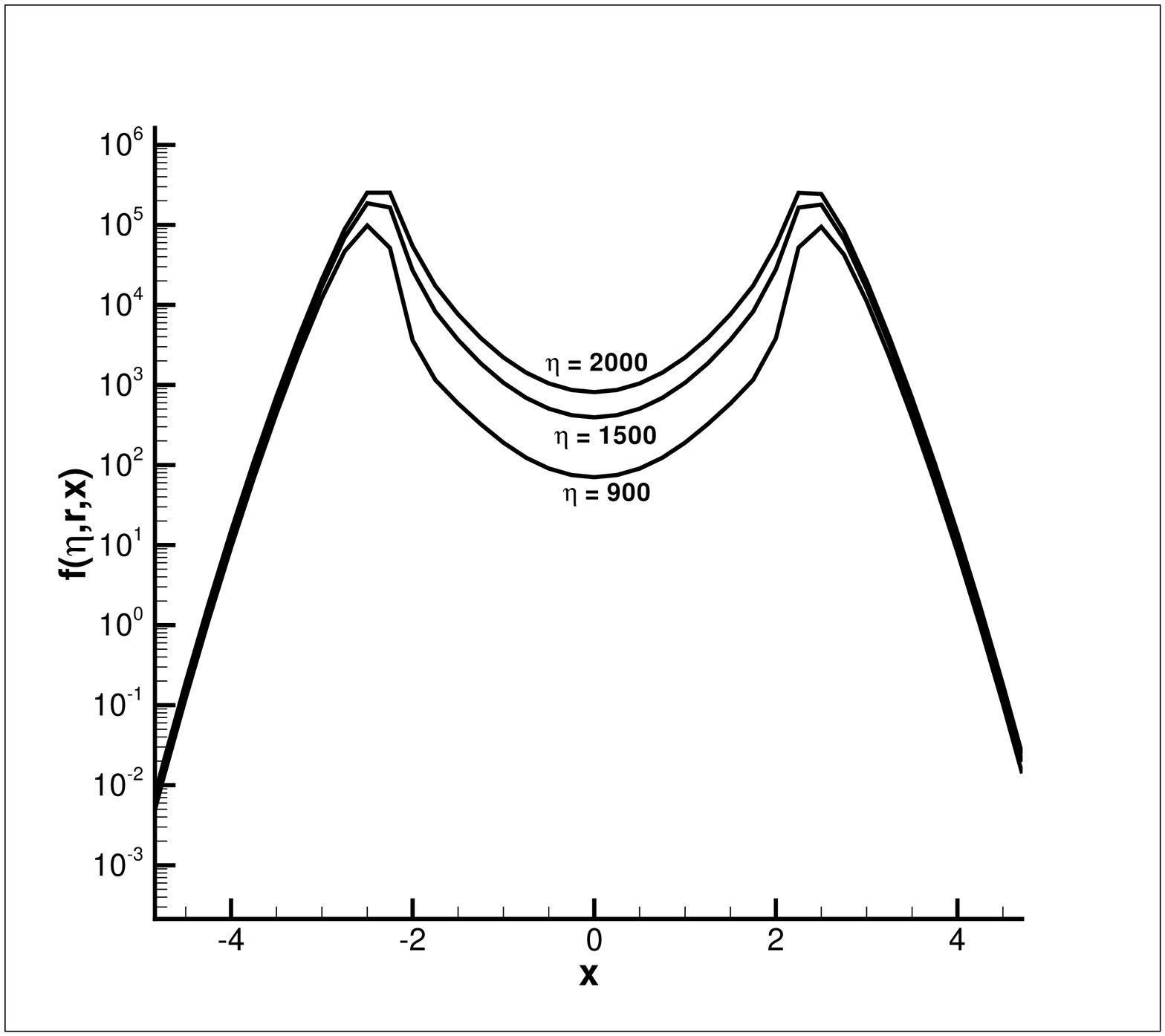}
\caption{The flux $f(\eta,r,x)$ of the solutions of eqs.(\ref{eq16}) and (\ref{eq17})
at $r=$ 50 (left), 100 (middle) and 500 (right). The frequency profile of the
source is $\phi_s(x)=(1/\sqrt{\pi/2})e^{-2x^2}$.
}
\label{fig5}
\end{figure}

More importantly, the amplitude of the peaks approaches its
saturation at about $\eta \simeq 200$ for $r=50$, $\eta\simeq 400$
for $r=100$, and $\eta\simeq 2,000$ for $r=500$. That is, the time $\eta$
needed for the spatial transfer over size $r$ is roughly proportional to
$r$. This is very different from the random walk relation $\eta \propto
r^2$, or the results shown in Figure \ref{fig3}.

It should be emphasized that the photons at the two peaks are not
only those with $x \simeq \pm (2-3)$ directly from the source $S_0
\phi_s(x)$, but also include the photons with frequency-shift from $x\simeq
0$ to $x \simeq \pm (2-3)$. This point can be shown by a source of
the profile with smaller width, say, $\phi_s(x)=(1/\sqrt{\pi/2})e^{-2x^2}$.
In Figure \ref{fig5}, we plot the results. For the  source of
$\phi_s(x)=(1/\sqrt{\pi/2})e^{-2x^2}$, the number of photons with
$x \simeq \pm (2-3)$ is much less than that of
the source $(1/\sqrt{\pi})e^{-x^2}$. We can see that all the features in Figure
\ref{fig5} are the same as those in Figure \ref{fig4}. The two peaks are
still located  at $x \simeq \pm (2-3)$. The time scale for approaching
saturation in Figure \ref{fig5} is also the same as that in Figure
\ref{fig4}.

Therefore, photons at the peaks of $x \simeq \pm (2-3)$ should come
from the frequency-shift from $x=0$ to $x\simeq \pm(2-3)$ by resonant
scattering. That is, resonant scattering provides a shortcut of the
spatial transfer: first to shift $x=0$ photons to $x\simeq \pm(2-3)$
photons, and then to speed  up spatial transfer. This
mechanism is the same as that for the escape of resonant photons from
opaque clouds (Osterbrock 1962; Harrington 1973; Avery \& House 1968; Adams
1972). It allows the Ly$\alpha$ photons to be able to transfer over to
the 21 cm regions in time $\eta$ proportional to its size
$r$.

\subsection{Resonant photon restoration and W-F coupling onset}

\begin{figure}[htb]
\centering
\includegraphics[width=5cm]{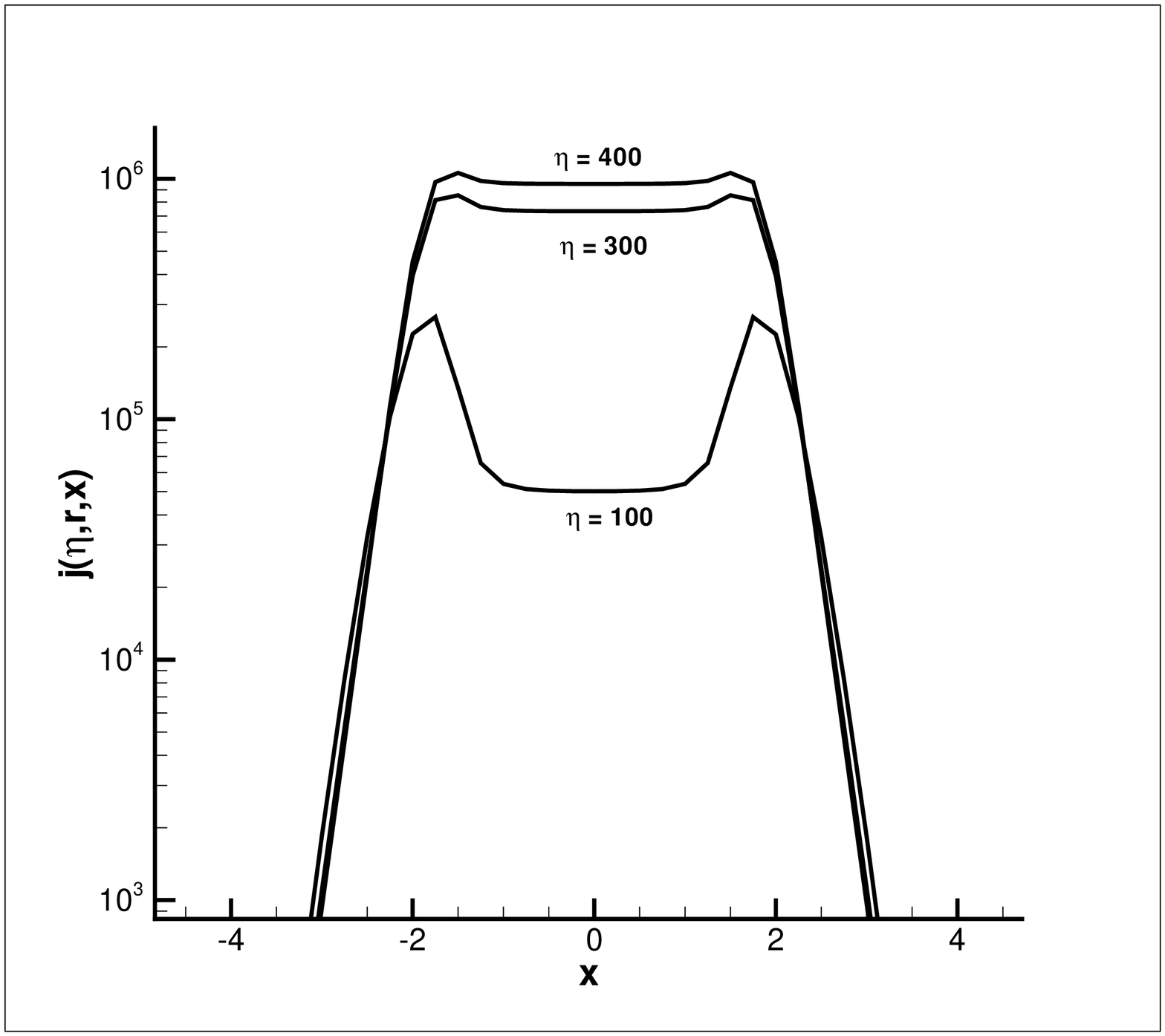}
\includegraphics[width=5cm]{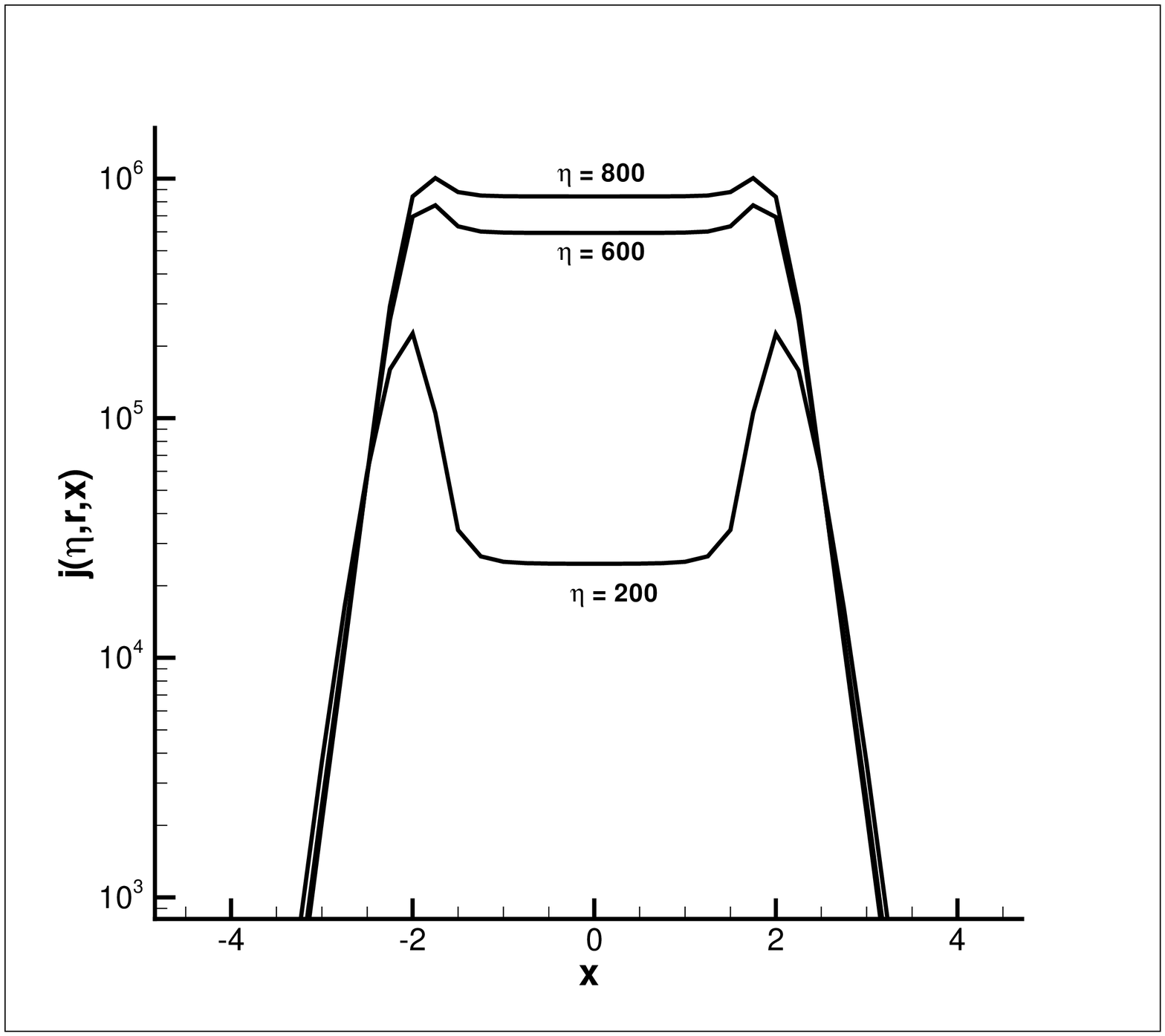}
\includegraphics[width=5cm]{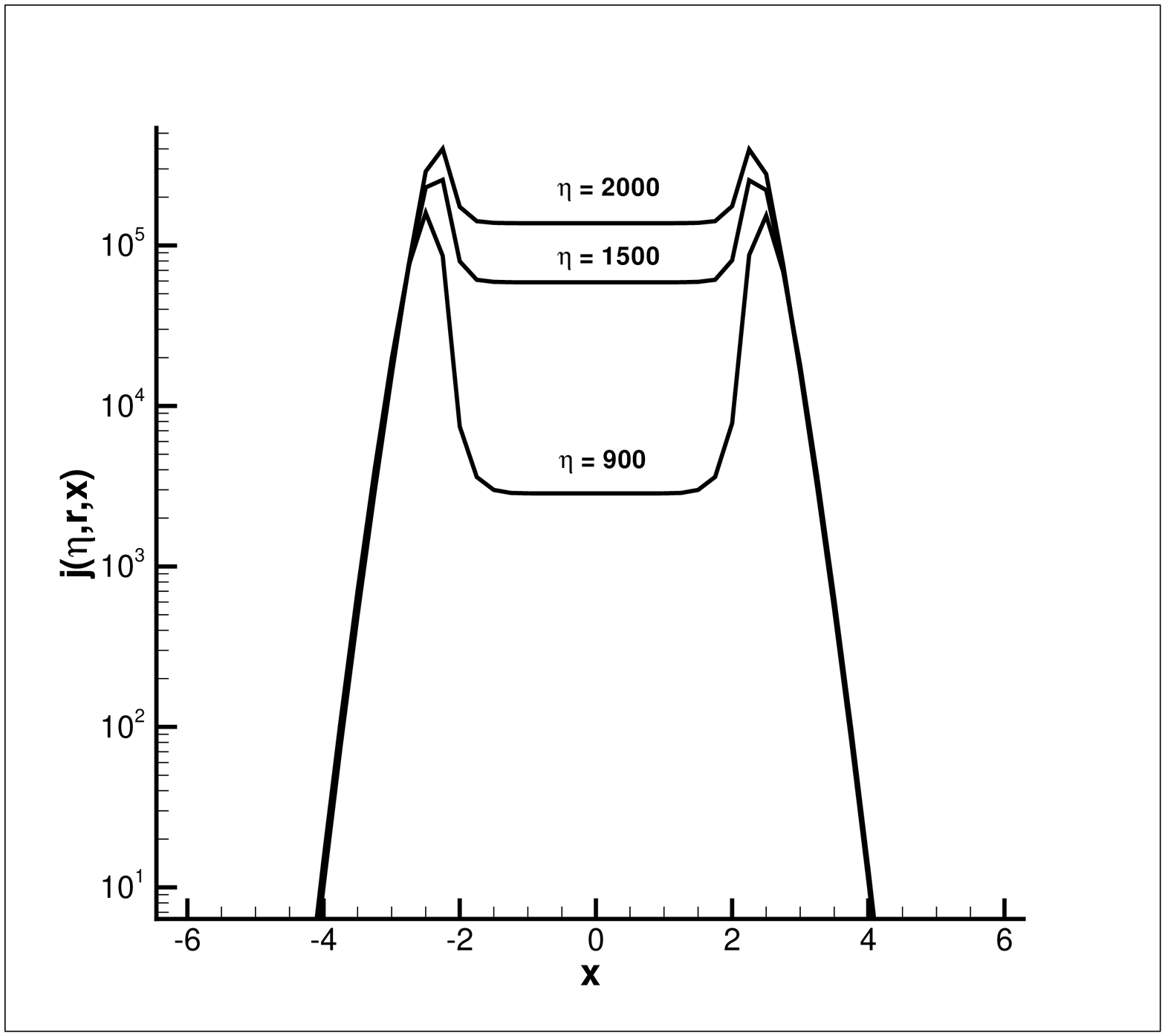}
\caption{The mean intensity $j(\eta,r,x)$ of eqs.(\ref{eq16}) and (\ref{eq17}) at $r=$ 50
(left), 100 (middle) and 500 (right).}
\label{fig6}
\end{figure}

We have mentioned in \S 4.2 that the flux $f(\eta,r,x)$ is lacking $\nu_0$
(or $x=0$) photons even when $f$ approaches its saturation. This is, obviously, not good for
the W-F coupling. However, we found that the situation may not be so if we consider
the solution of the mean specific intensity $j$. Figure \ref{fig6} presents the solution
$j(\eta,r,x)$ of equations (\ref{eq16}) and (\ref{eq17}) at $r=$ $50$, $100$ and $500$.
The parameters $b=0$ and $\gamma=10^{-5}$ are the same as the solutions in
Figure \ref{fig4}. The results are given in Figure \ref{fig6}.

When the time $\eta$ is small, $j(\eta,r,x)$ is similar to $f(\eta,r,x)$, having a valley
around $x=0$ and two peaks at  $x\simeq \pm (2-3)$. However, different from the flux
$f(\eta,r,x)$, the amplitude of $j(\eta,r,x)$ around $x=0$
is quickly increasing. In the saturated state it is about the same as the peaks. That is,
although the flux always has a valley around $x=0$, the $\nu_0$ photons are quickly
restored in the mean intensity. The time scale of approaching its saturated state is
also proportional to $r$, not $r^2$. Therefore, the restoration of resonant photons is due to
the resonant scattering ``bounce back'' (\S 3.3), which pushes photons with frequency
$x\simeq \pm( 2-3)$ back to $x \simeq 0$.

\begin{figure}[htb]
\center
\includegraphics[width=6cm]{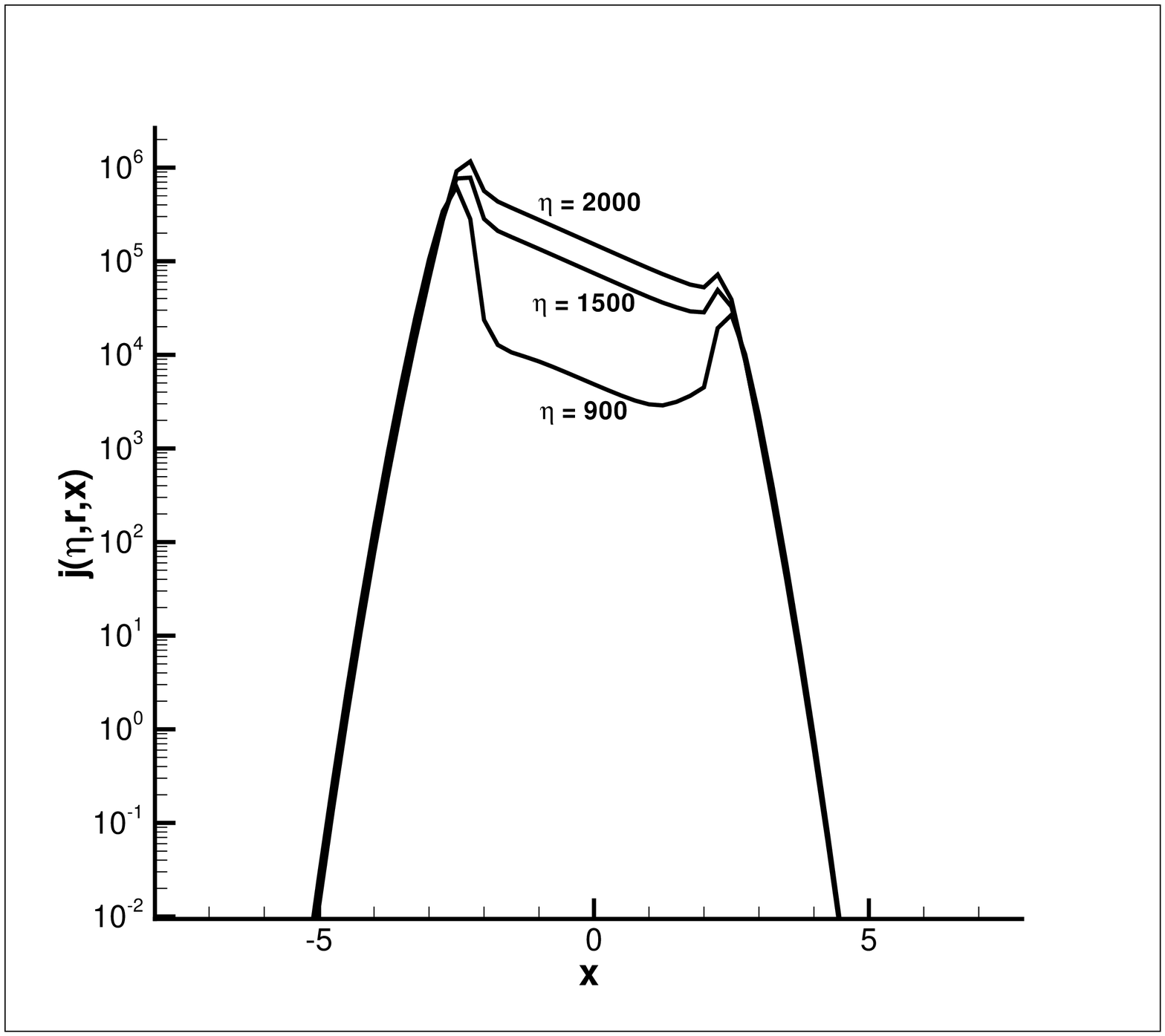}
\includegraphics[width=6cm]{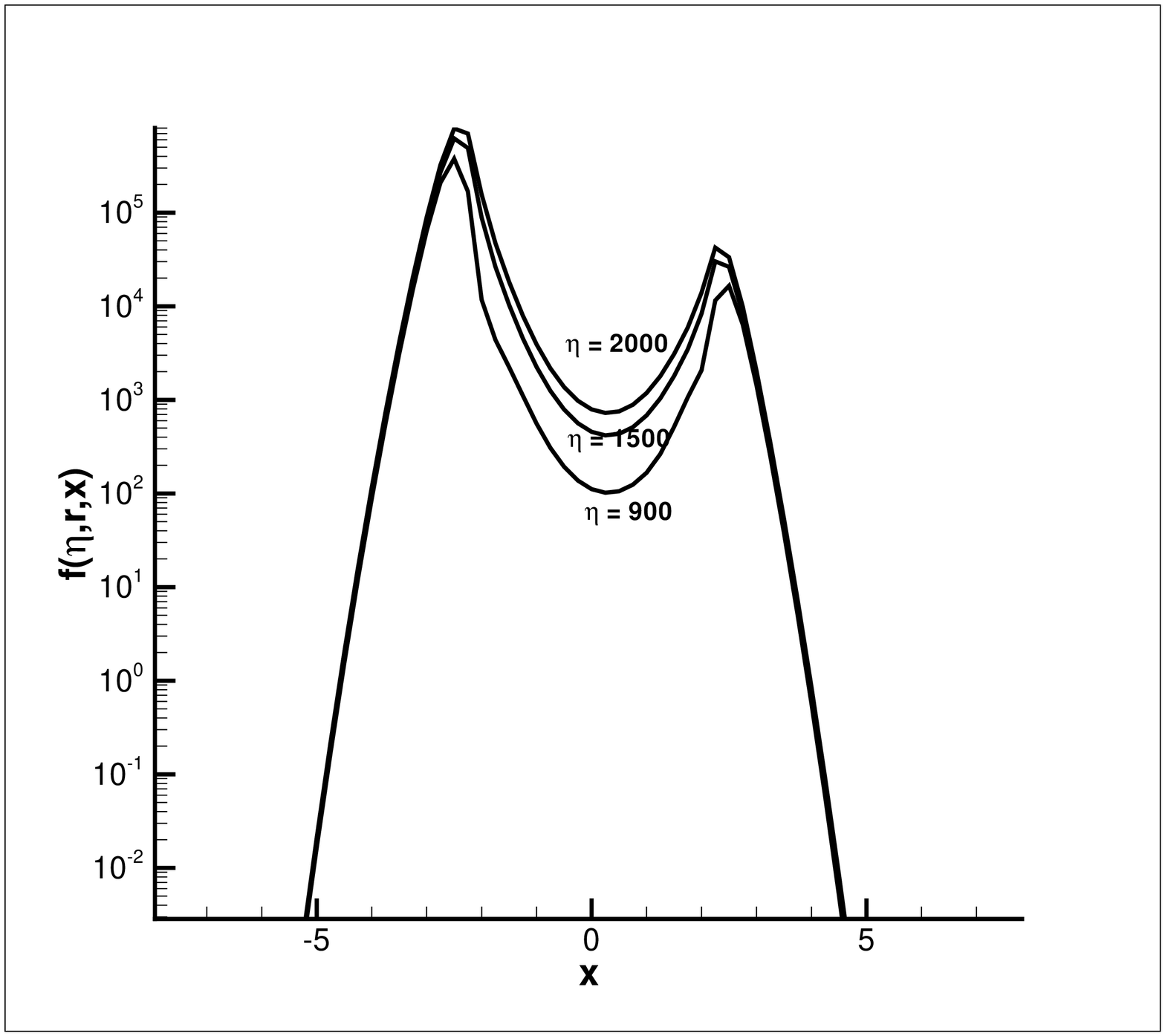}
\caption{The mean intensity $j(\eta,r,x)$ (left) and
flux $f(\eta, r,x)$ (right) of the solution of eqs.(\ref{eq16}) and (\ref{eq17}) at
$r=$ 500. The recoil parameter is taken to be $b=0.3$.}
\label{fig7}
\end{figure}

The shape of $j(\eta,r,x)$ around $x=0$ is a flat plateau, which is similar to that
in Figure \ref{fig1}. As have been shown in PaperI, the flat plateau of $j(\eta,r,x)$
at $b=0$ will become the local Boltzmann distribution if the recoil is considered.
We can expect that the flat plateau of Figure \ref{fig6} will also show a local
Boltzmann distribution if $b\neq 0$. We calculate the solutions $j(\eta,r,x)$ and
$f(\eta,r,x)$ of eqs. (\ref{eq16}) and (\ref{eq17}) at $r=500$ with the same
parameters as those in Figure 6, but with $b=0.3$. We use a large $b$, because it
is easier to see the slope of the local Boltzmann distribution. The results are
given in Figure \ref{fig7}.

Figure \ref{fig7} clearly shows a local Boltzmann distribution within the range $|x|\leq 2$
as
\begin{equation}
\label{eq23}
j(\eta,r,x)\simeq j(\eta,r,0)e^{-2bx}= j(\eta,r,0)e^{-h(\nu-\nu_0)/k_BT}.
\end{equation}
We find the slope to be $\ln j(\eta, 0) -\ln j(\eta, 1) = 0.59$, which is
well consistent with $2b=0.6$. Figure \ref{fig7} shows that at $r=500$, the
onset of the W-F coupling can occur as early as $\eta=900$, but the
amplitude of the local Boltzmann distribution at that time is much lower than its
saturated value by a factor of 10$^2$. The amplitude of the local
Boltzmann distribution is substantially increasing with time.
Therefore, the ``bounce back'' mechanism keeps the W-F coupling to
work with a timescale $\eta$ larger than a few hundreds, i.e. a
few hundred collisions. This result is the same as that in PaperI.

The solution of the flux $f(\eta, r,x)$ in Figure \ref{fig7} is not very different from
that in Figure \ref{fig4}.  The only difference between the two figures is
that the former is asymmetric with respect to $x=0$,
i.e. the peak at $x<0$ is stronger than that of $x>0$, while the latter is
symmetric. This is simply due to the recoil of $b\neq 0$ leading more photons
to move to $x<0$. Neither  flat plateau nor local Boltzmann distribution
is shown in the flux $f(\eta, r,x)$. There is always a valley around $\nu_0$
even when $f(\eta, r,x)$ is in its saturated state. It once again indicates
that the flux is dominated by photons of $x\simeq \pm( 2-3)$.

\begin{figure}[htb]
\center
\includegraphics[width=7cm]{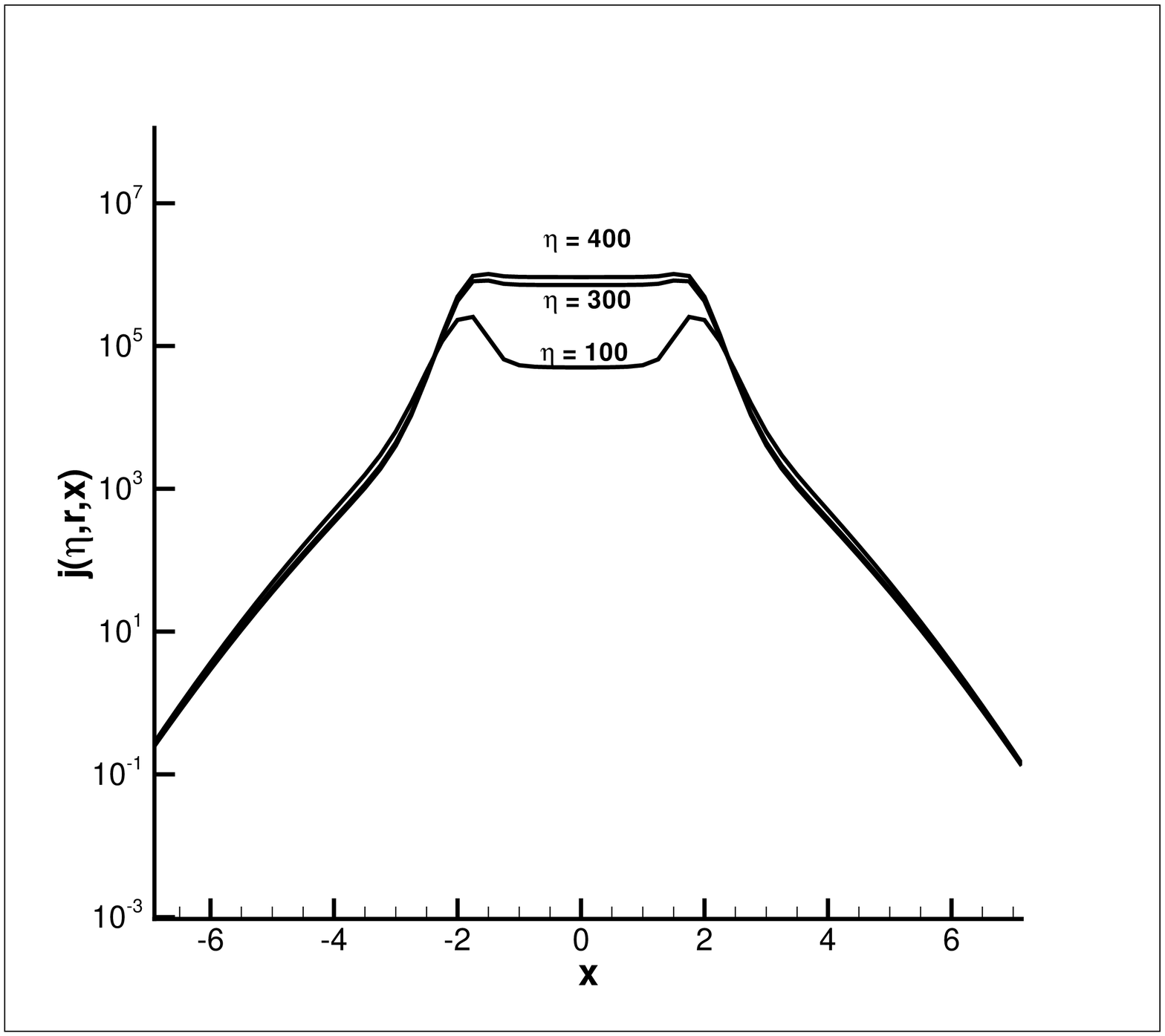}
\includegraphics[width=7cm]{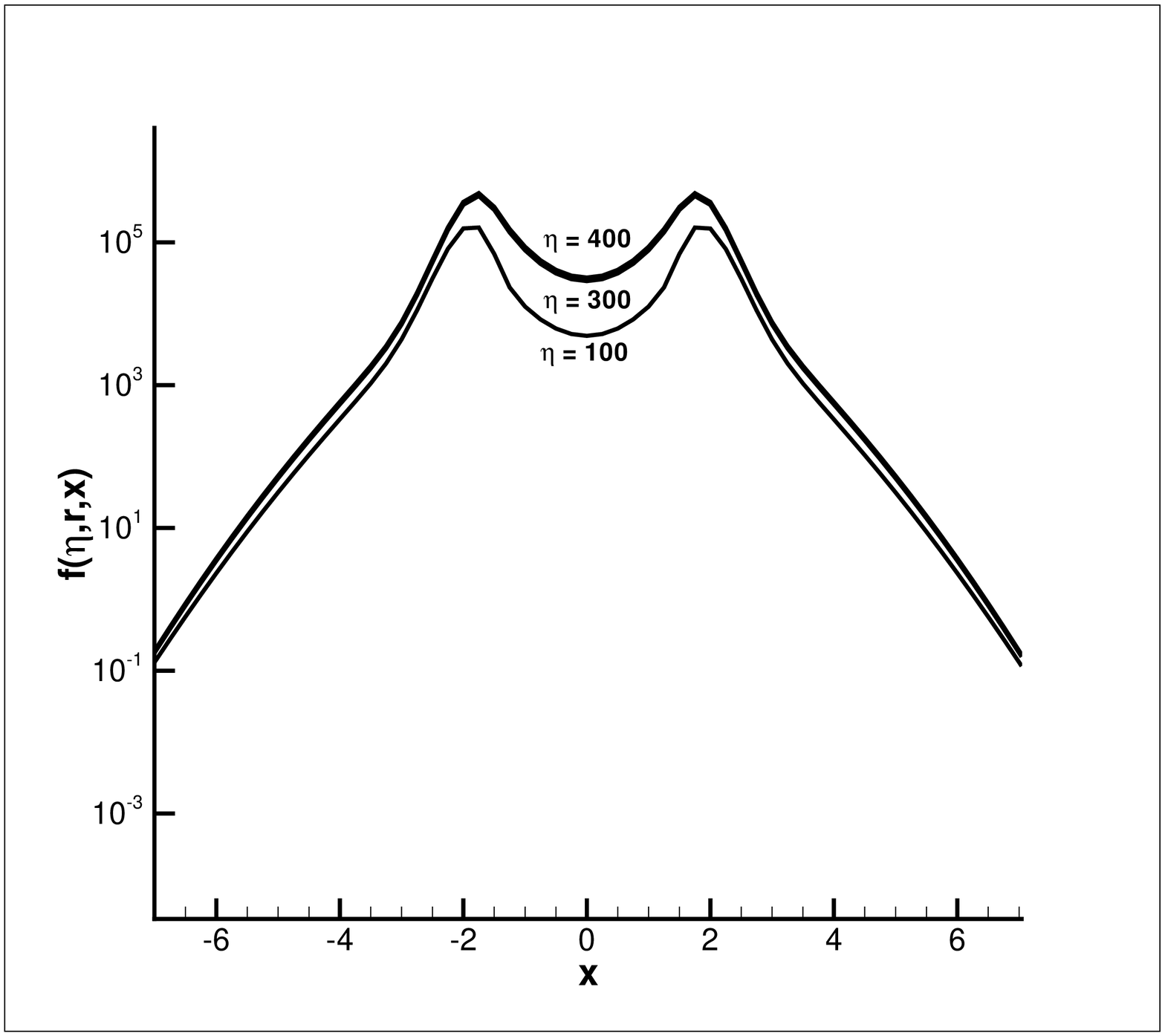}
\caption{The mean intensity $j(\eta,r,x)$ (left) and flux $f(\eta,
r,x)$ (right) of the solution of eqs.(\ref{eq16}) and (\ref{eq17})
with the Voigt profile eq.(\ref{fig4}) and re-distribution function eq.(\ref{fig7}).
The parameters are taken to be $r=$ 50 and $a=10^{-2}$.}
\label{fig8}
\end{figure}

The two components $f$ and $j$ of the Eddington approximation are
effective in revealing the functions of the resonant scattering. The
flux $f(\eta,r,x)$ describes the photons in transit. It shows that
the spatial transfer within opaque media is mainly via photons with
frequency shifted to $x\simeq \pm (2-3)$, which are easy for
escaping. The mean intensity $j(\eta,r,x)$ describes the restoration
of $x=0$ photons and the onset of the W-F coupling. Therefore,
$f(\eta,r,x)$ shows a deep valley around $x=0$ (or $\nu_0$), while
$j(\eta,r,x)$ shows a plateau.

In Figure \ref{fig8}, we present the the solutions of mean intensity $j(\eta,r,x)$ and flux
$f(\eta, r,x)$ given by eqs.(\ref{eq16}) and (\ref{eq17})
with the Voigt profile eq.(\ref{fig4}) and re-distribution function eq.(\ref{fig7}). As expected,
Figure \ref{fig8} has the Lorentz wings. However, in the center part $|x|\leq 3$,
Figure \ref{fig8} shows the same features as Figures \ref{fig4} and \ref{fig6}.
That is, the mechanism of
``escape via shortcut'' plus ``bounce back'', which mainly relies on photons
with $|x| < 3$, still works well. The Lorentz wing only leads to long tails
in the profiles of $j$ and $f$ in the range $|x|>3$, and the wings have very
low amplitudes.

\subsection{Effect of injected photons}

\begin{figure}[htb]
\begin{center}
\includegraphics[width=5cm]{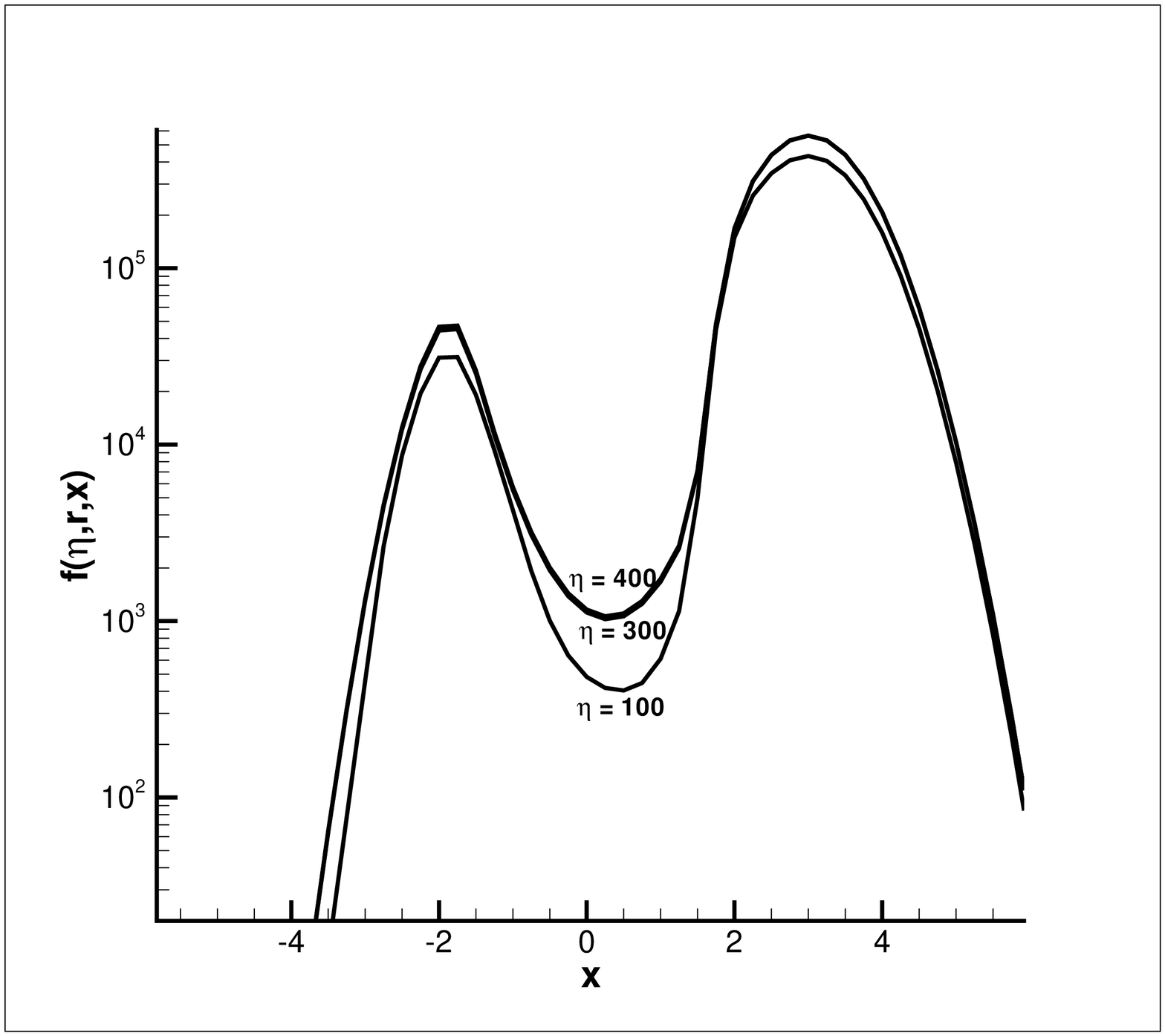}
\includegraphics[width=5cm]{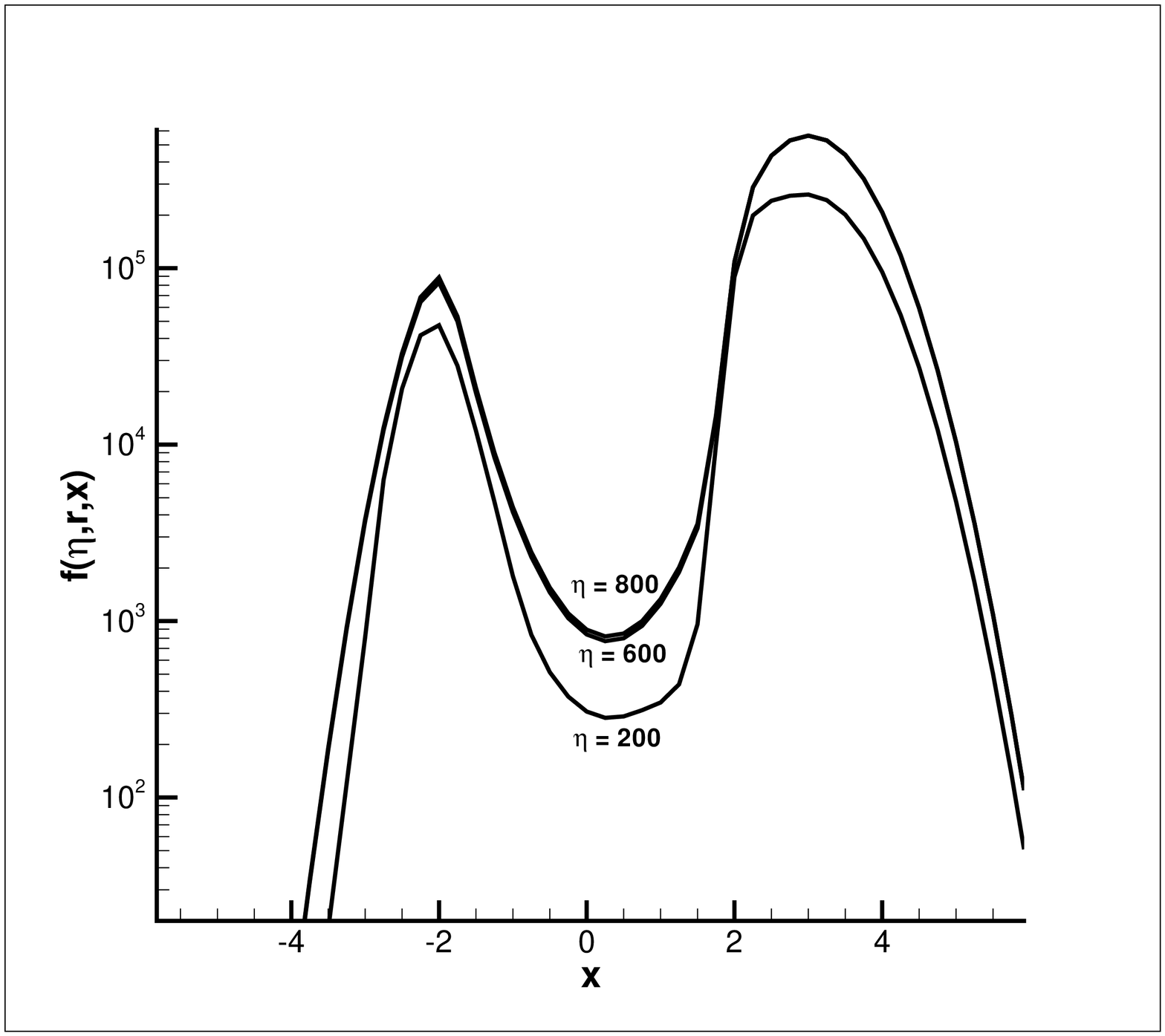}
\includegraphics[width=5cm]{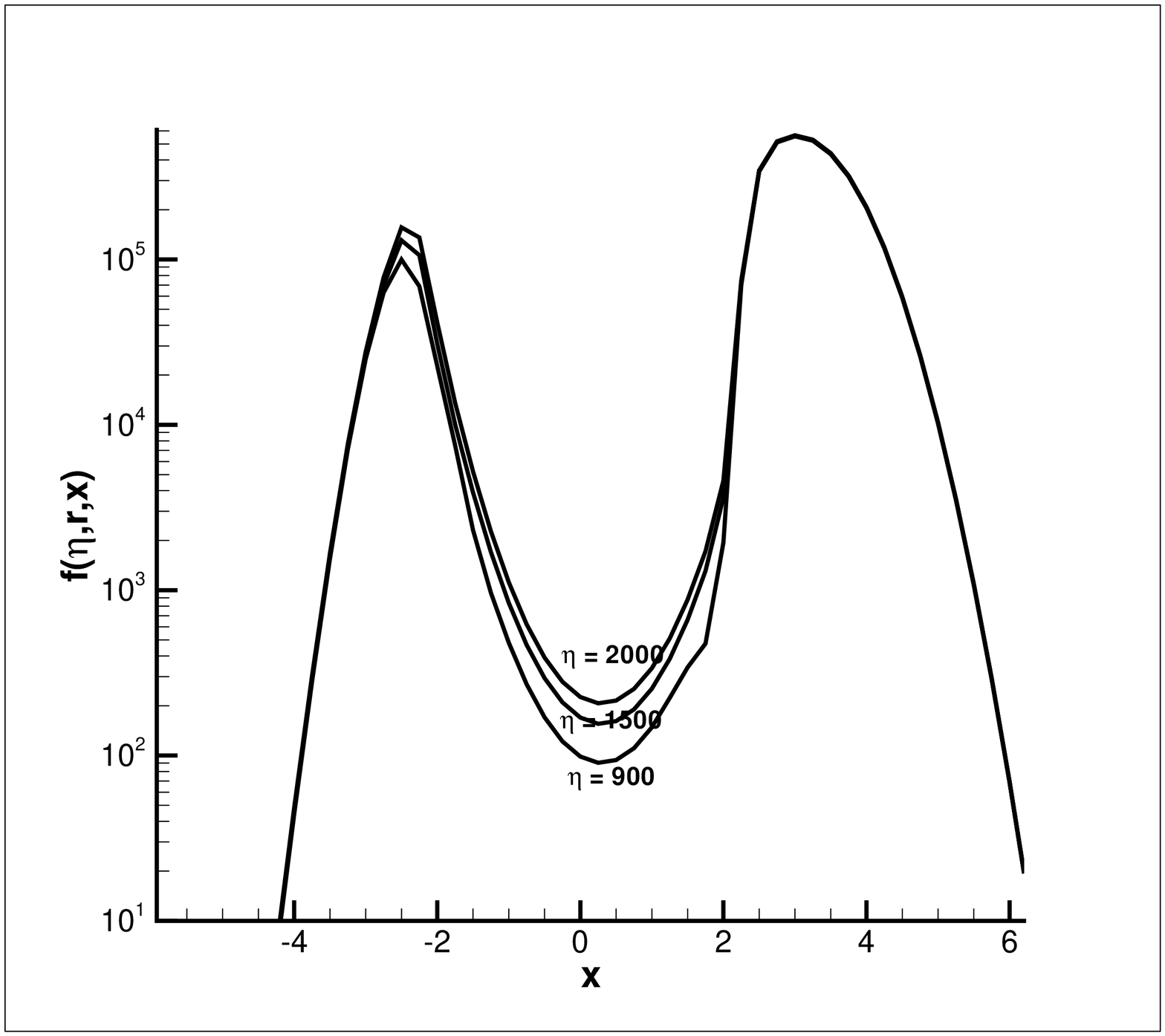}
\end{center}
\begin{center}
\includegraphics[width=5.0cm]{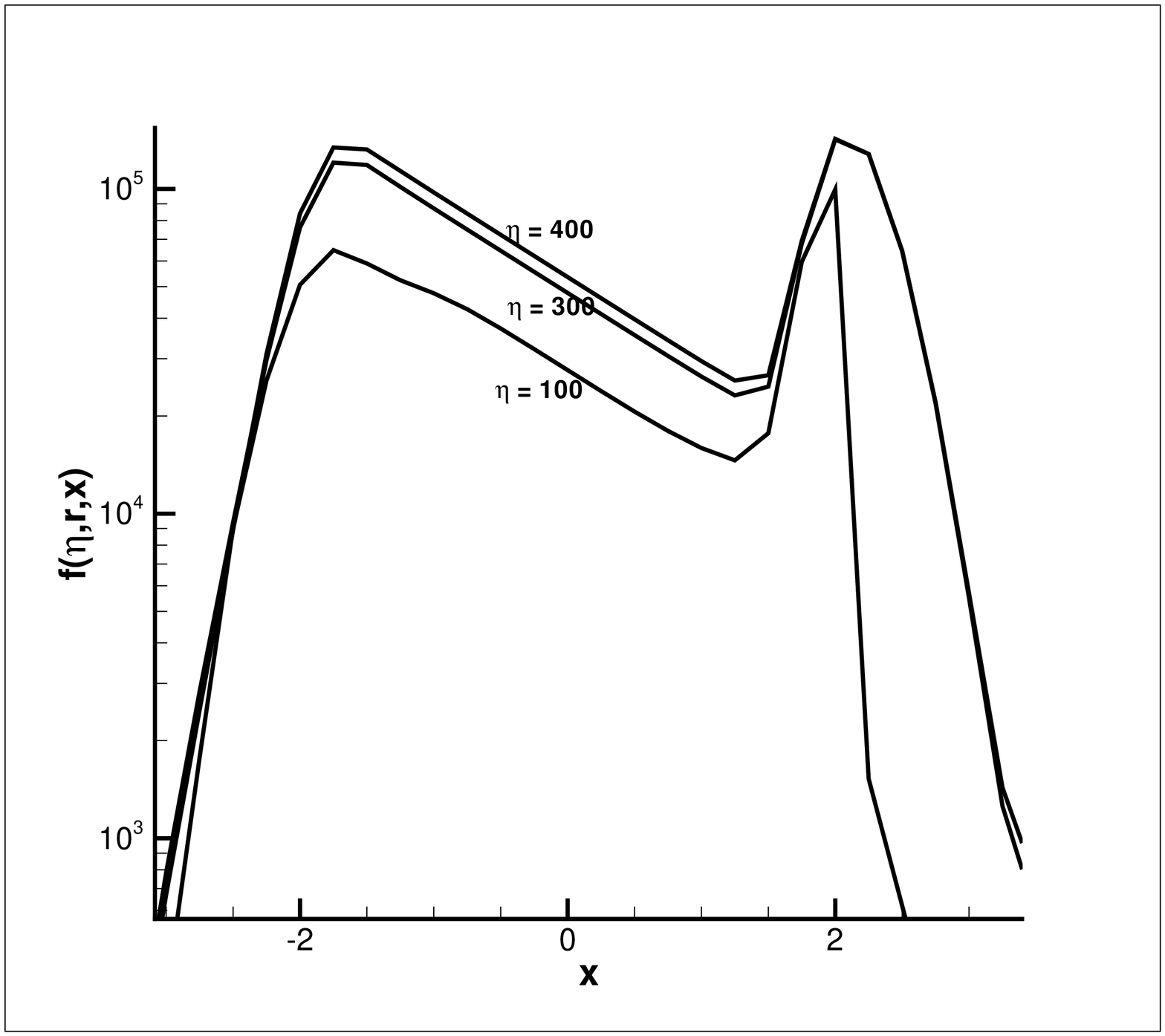}
\includegraphics[width=5.0cm]{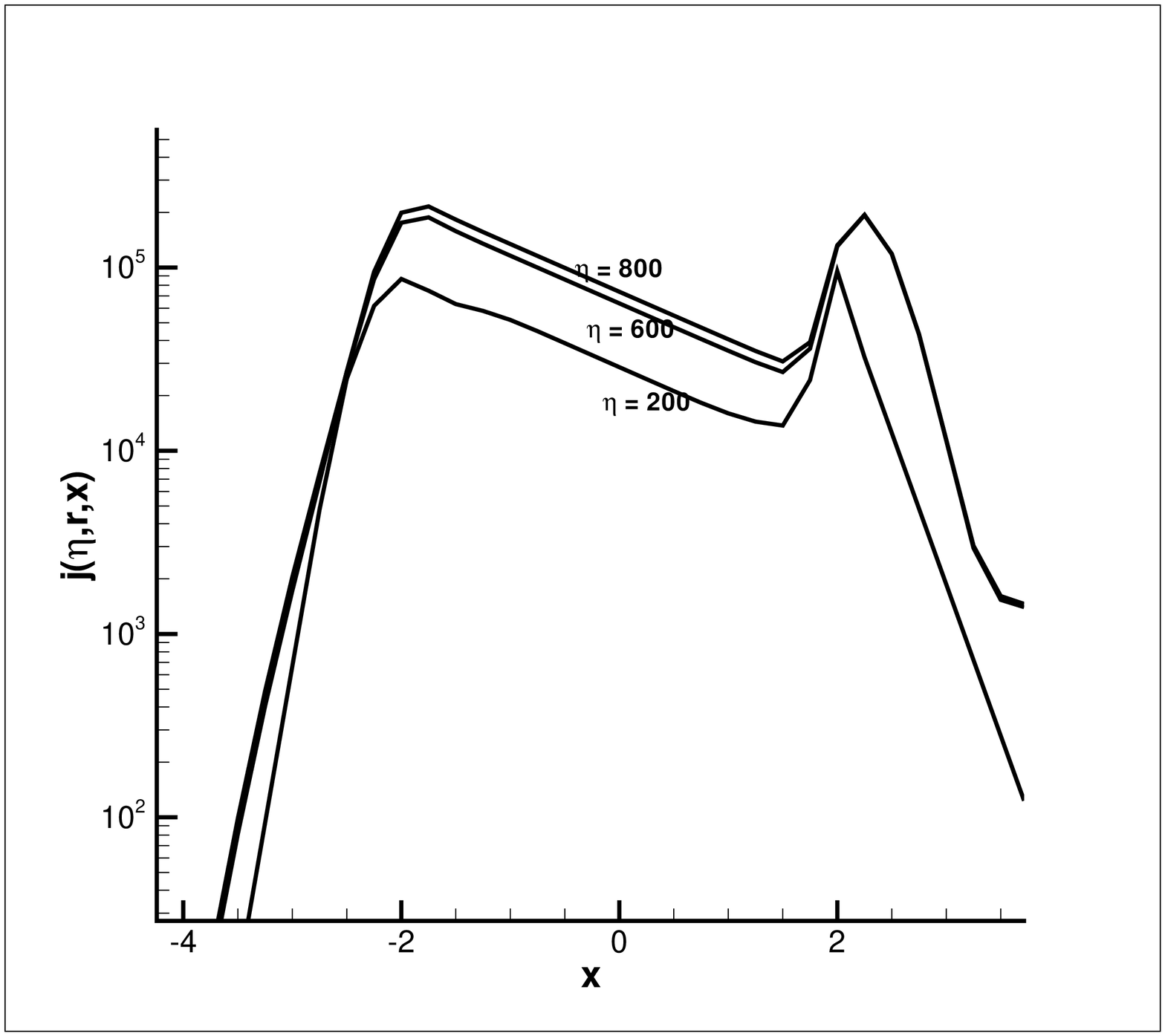}
\includegraphics[width=5.0cm]{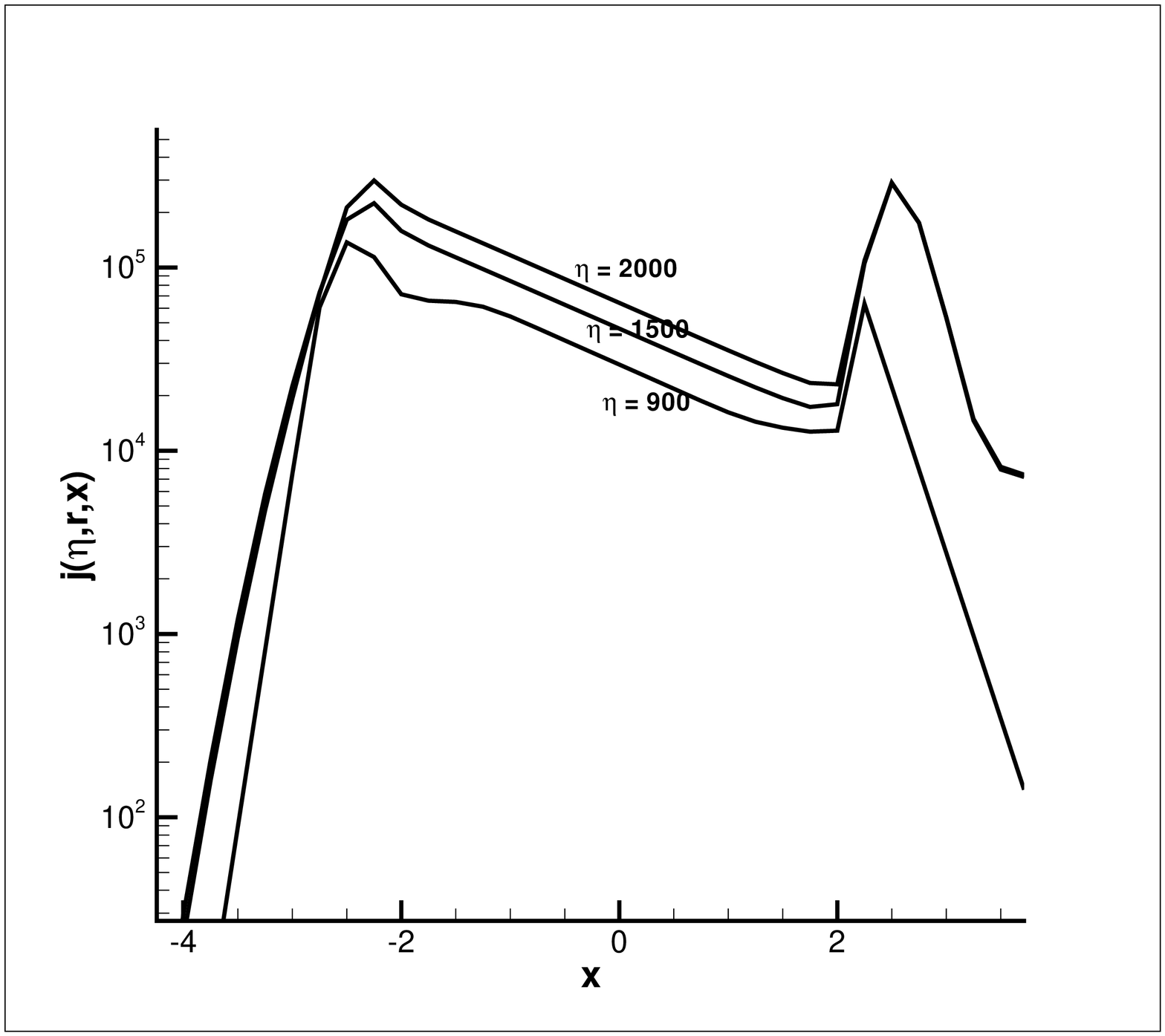}
\end{center}
\caption{The mean intensity $j(\eta,r,x)$ and flux $f(\eta,r,x)$ of
eqs.(\ref{eq16}) and (\ref{eq17}) at $r=$ 50 (left), 100 (middle)
and 500 (right). parameter $b=0.3$. The frequency profile of the source is
$\phi_s(x)=\phi_g(x-3)$.}
\label{fig9}
\end{figure}

Hubble redshift is another mechanism to produce photons with
frequency $\sim \nu_0$, which also have the problems of the
spatial transfer and the W-F coupling. Since the 21 cm region
basically is optical thin for photons with $x>3$, we first model the
cosmic redshift by a photon source with the profile
$\phi_s(x)=\phi_g(x-3)$. The solutions of $f(\eta,r,x)$ and
$j(\eta,r,x)$ with the same parameters as in Figure \ref{fig7} are
shown in Figure \ref{fig9}.

Although the injected photon has a peak at $x=3$, the flux $f(\eta,r,x)$ in Figure
\ref{fig9} still shows two peaks at $x\simeq \pm( 2-3)$. The peak at $x\simeq (2-3)$
is higher than that at $-x\simeq (2-3)$. It is simply because the photon source is
at $x=3$. The peak at $x<0$ is also much higher than $f(\eta,r,x)$ at $x=0$. That is,
even though the source photon is at $x=3$, the spatial transfer is still dominated
by photons of both $x
\simeq (2-3)$ and $x \simeq -(2-3)$. The mean intensity $j(\eta,r,x)$ shows once again
a perfect local Boltzmann distribution with slope $2b$ in the range $-2 <x<2$
(eq.(\ref{eq23})). Any photons injected by the redshift into the 21 cm region
will quickly join the W-F coupling by the bounce back mechanism.

\begin{figure}[htb]
\begin{center}
\includegraphics[width=7cm]{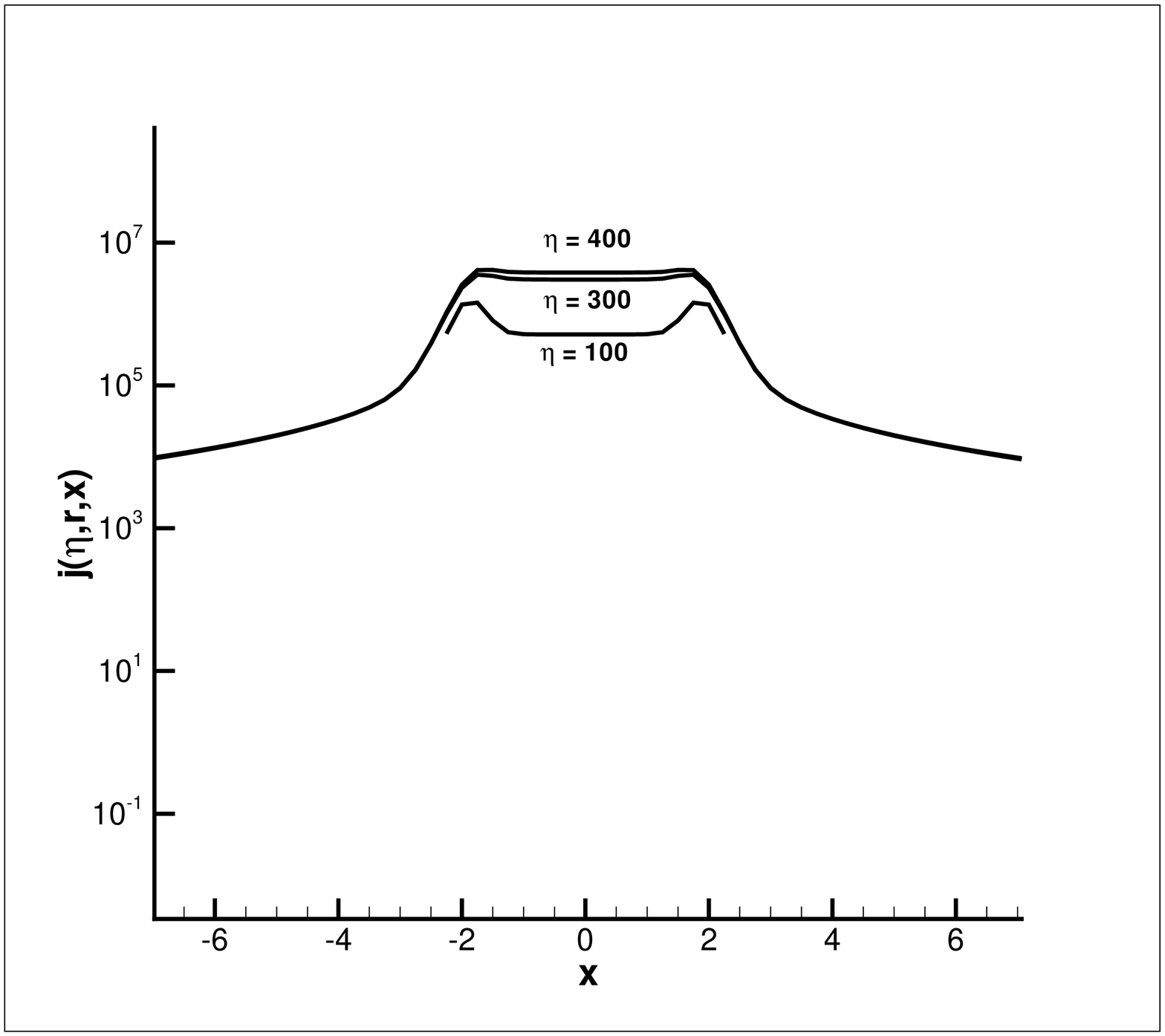}
\includegraphics[width=7cm]{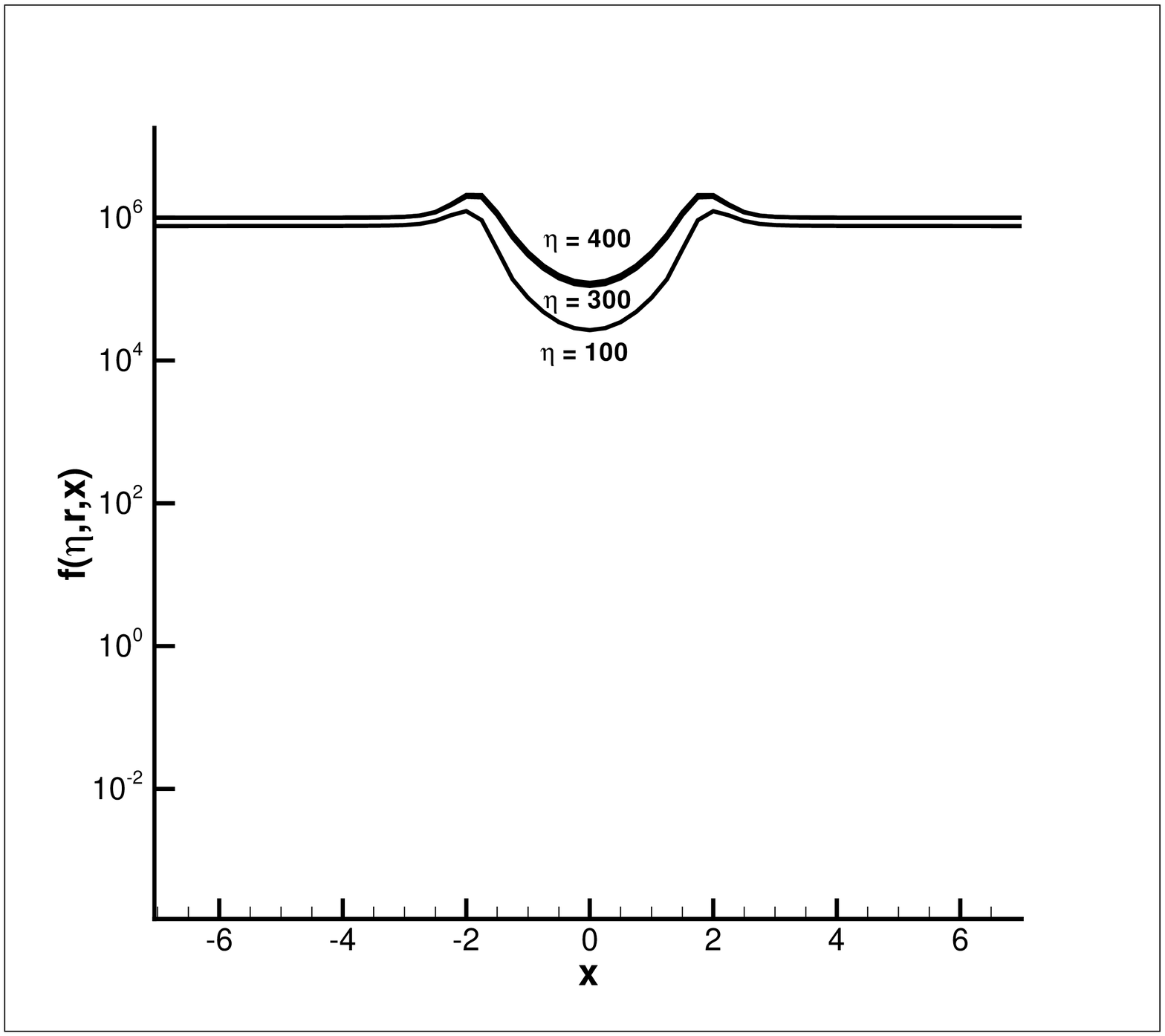}
\end{center}
\caption{The mean intensity $j(\eta,r,x)$ and flux $f(\eta,r,x)$ of
eqs.(\ref{eq16}) and (\ref{eq17}) with the Voigt profile eq.(\ref{fig4}) and re-distribution
function eq.(\ref{fig7}). The frequency profile of the source is flat, i.e. $\phi_s(x)$
is equal to $S_0$ within
$|x|\leq 10$, and to zero at $|x|> 10$.}
\label{fig10}
\end{figure}

We now model the source with a continuous frequency spectrum: $\phi_s(x)=S_0$ within
$|x|\leq 10$, and $\phi_s(x)=0$ at $|x| > 10$. The solutions of the mean intensity
$j(\eta,r,x)$ and flux $f(\eta,r,x)$ of eqs.(\ref{eq16}) and (\ref{eq17}) are shown
in Figure \ref{fig10}, in which the other parameters are the same as those in
Figure \ref{fig8}.  We can see that
in the center part $|x|\leq 3$, $j(\eta,r,x)$ and $f(\eta,r,x)$ have the same features
as in Figure \ref{fig8}.  This result is expected, as the center part has contributions
from the photons
redshifted to $|x|\leq 3$ from $x >3$. Figure \ref{fig9} shows that the source of
$\phi_g(x-3)$ does not change the mechanism of ``escape via shortcut'' plus ``bounce back'',
and therefore, the source with a continuous frequency spectrum will keep these features
as well.

\subsection{Effect of the Hubble redshift}

\begin{figure}[htb]
\begin{center}
\includegraphics[width=5.0cm]{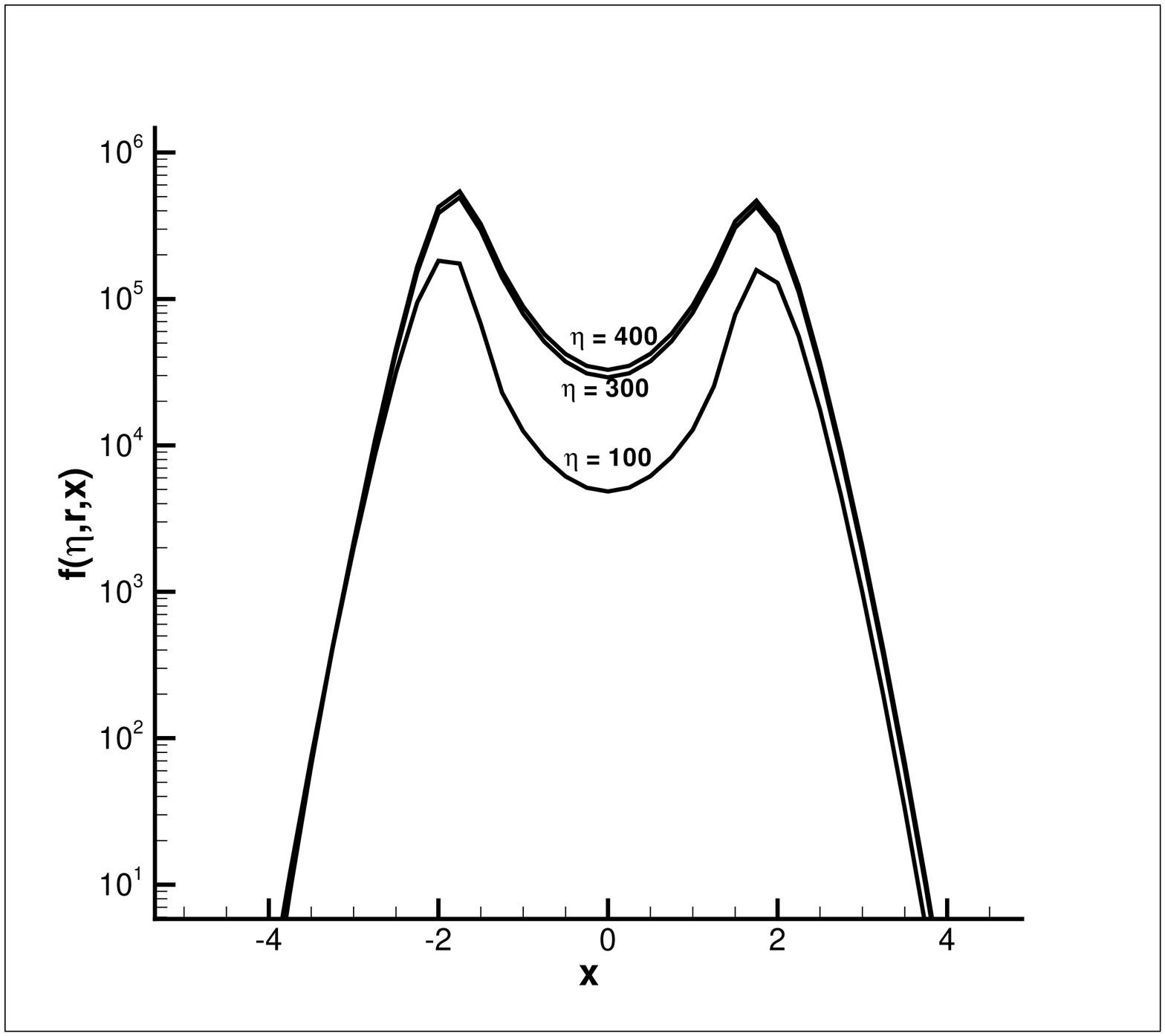}
\includegraphics[width=5.0cm]{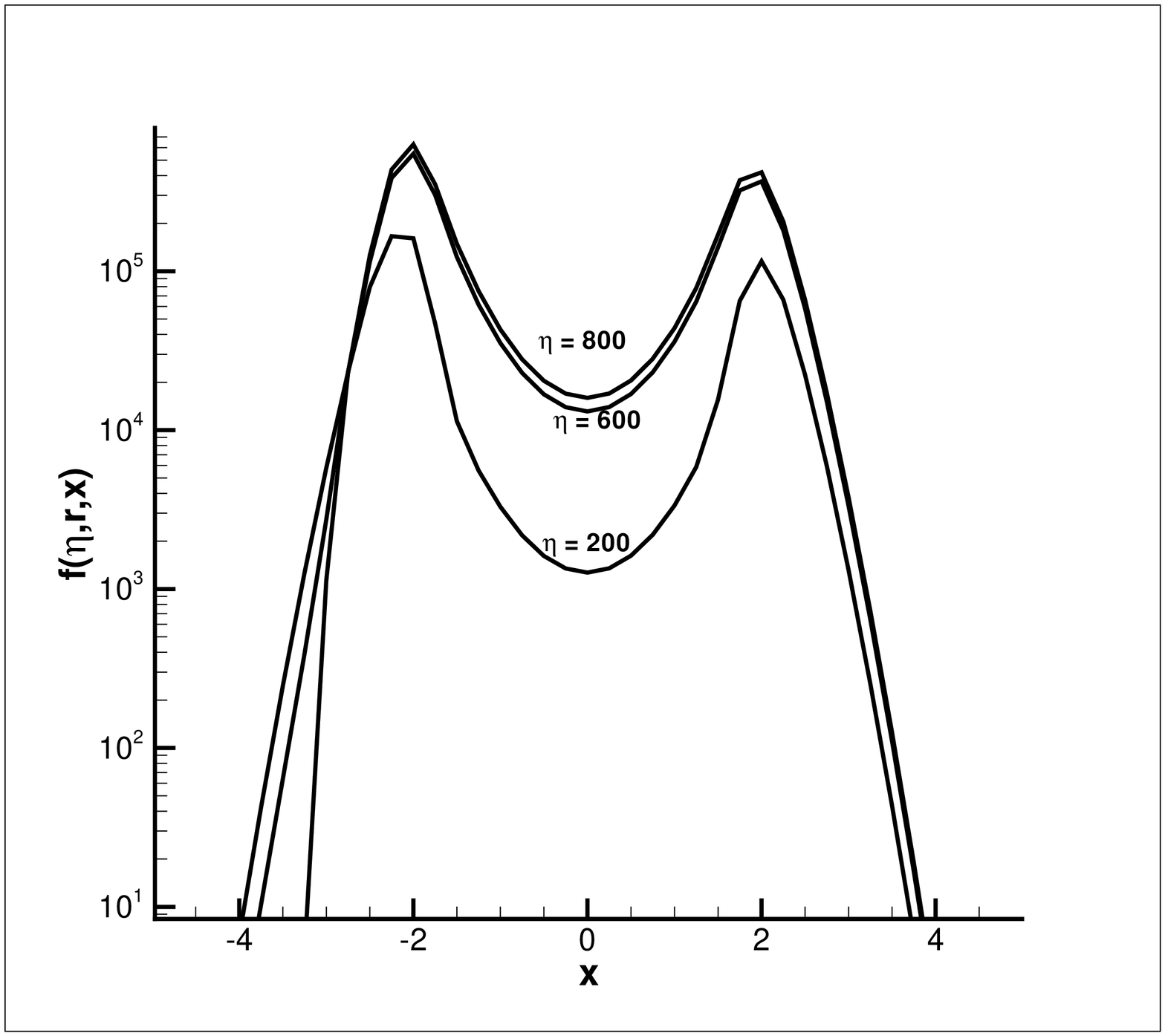}
\includegraphics[width=5.0cm]{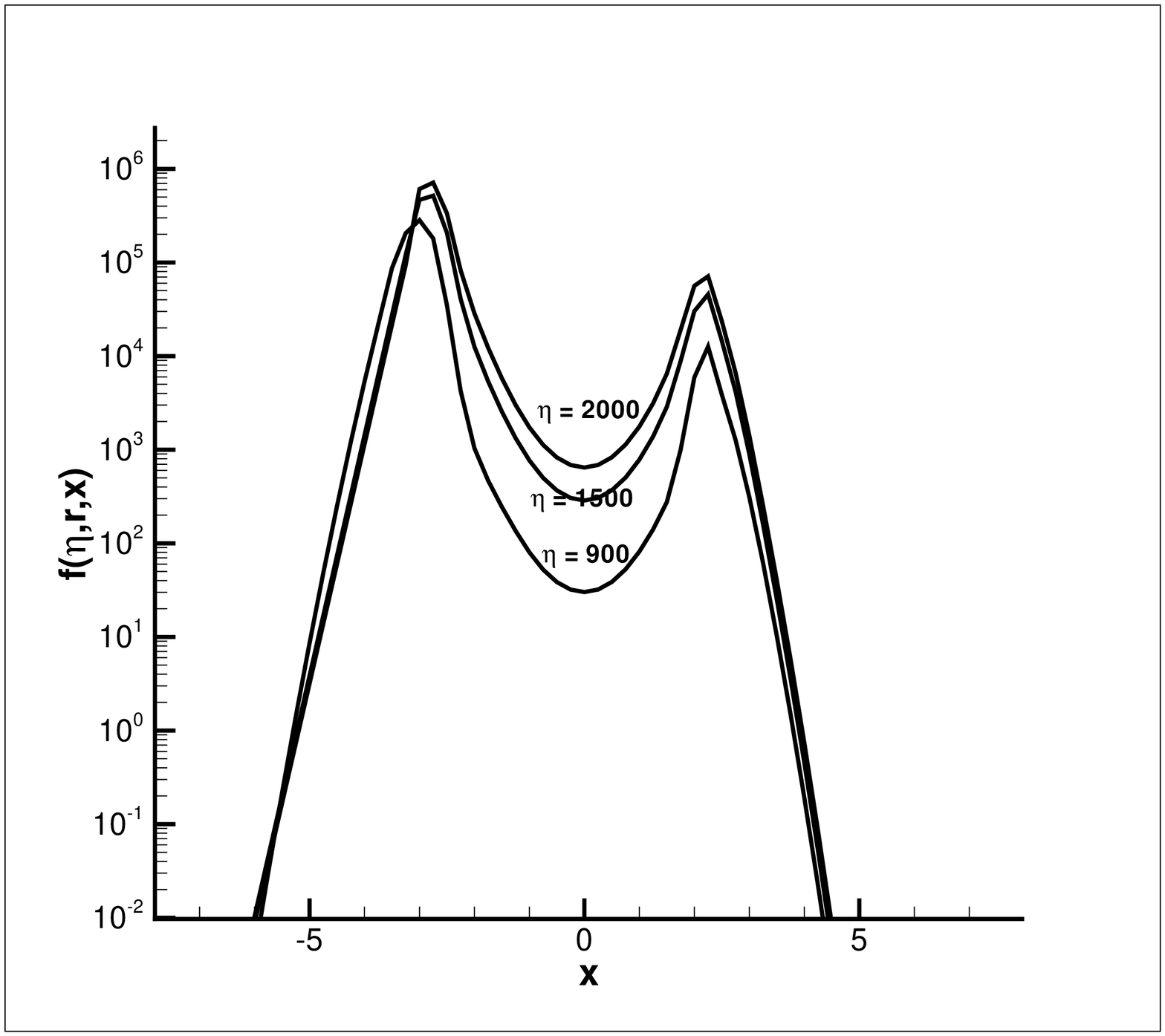}
\end{center}
\begin{center}
\includegraphics[width=5.0cm]{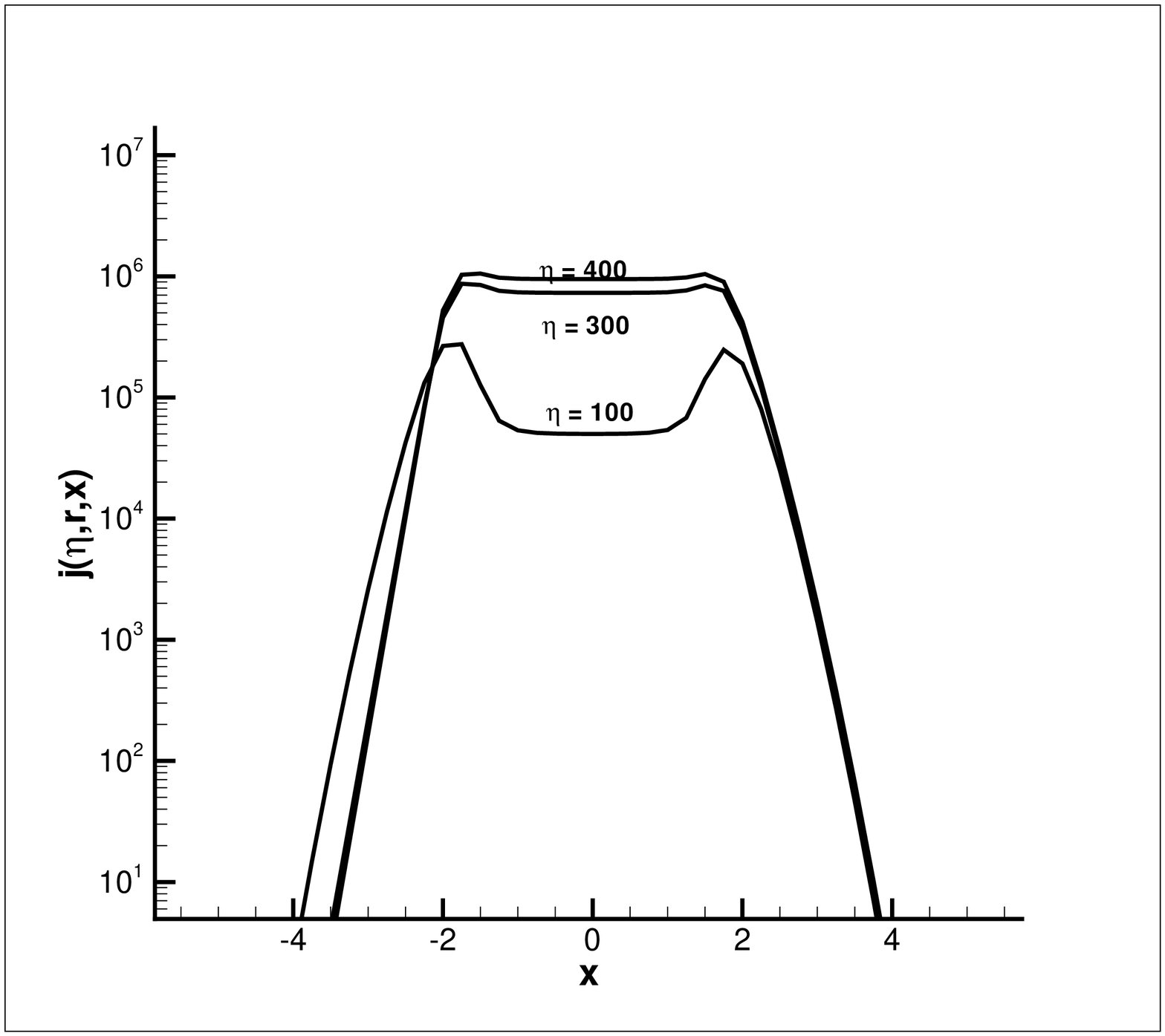}
\includegraphics[width=5.0cm]{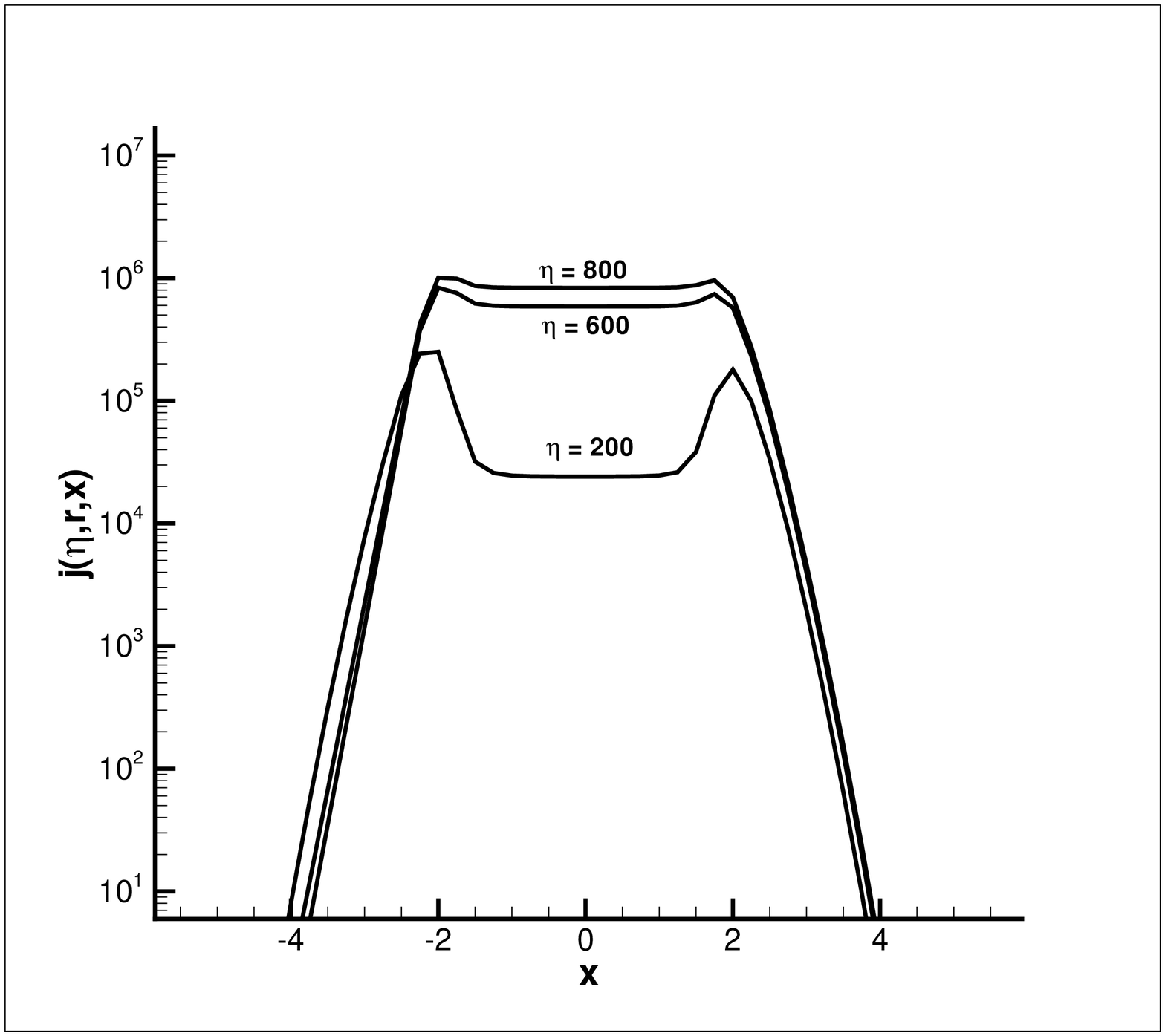}
\includegraphics[width=5.0cm]{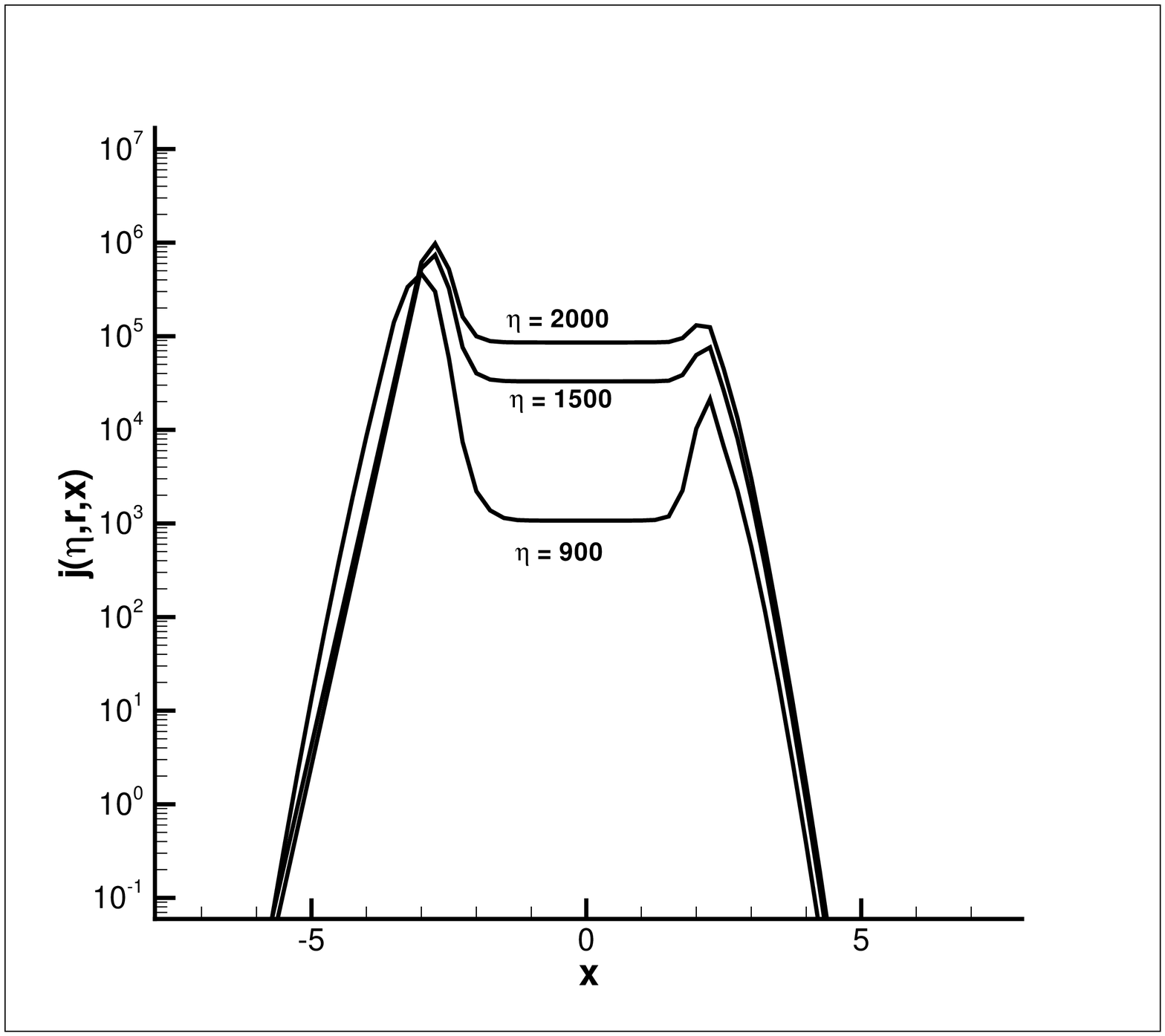}
\end{center}
\caption{The solutions $j(\eta,r,x)$ and $f(\eta,r,x)$ of eqs.(\ref{eq16})
and (\ref{eq17}) at $r=$ 50 (left), 100 (middle) and
500 (right). $\gamma=10^{-3}$. The parameters are the same as
those in Figures \ref{fig4} and \ref{fig6}, except that $\gamma=10^{-3}$.}
\label{fig11}
\end{figure}

Although the cosmic expansion is considered in all the above-mentioned
solutions, the effect of cosmic expansion seems to be negligible. It is
because the optical depth is large and the parameter $\gamma$ is very
small. The number of resonant scattering within the Hubble time is very
large. The cosmic expansion is too small in one free flight time.
The ``bounce back'' is dominant.

When $f_{\rm HI}$ is smaller, optical depth of the $\nu_0$ photons is
smaller, and $\gamma$ is larger, the effect of the Hubble redshift would
appear. Figure \ref{fig11} presents the solution $f(\eta,r,x)$ and
$j(\eta,r,x)$ of eqs.(\ref{eq16}) and (\ref{eq17})  with the same
parameters as those in Figures \ref{fig4} and \ref{fig6},
except that the parameter $\gamma=10^{-3}$ is large. Both $j(\eta,r,x)$ and $f(\eta,r,x)$
show the same features as in Figures \ref{fig4} and \ref{fig6}. The
Hubble redshift makes the profile to be asymmetric with
respect to $x=0$.  The red wing is stronger than the blue wing.
$j(\eta,r,x)$ still shows a flat plateau in the range $-2<x<2$.
Therefore, the W-F coupling will work when $\gamma=10^{-3}$,
corresponding to $f_{\rm HI}\simeq 10^{-3}$.

\section{An example of time-dependent effect of W-F coupling on 21 cm signals}

\begin{figure}[htb]
\begin{center}
\includegraphics[height=9cm]{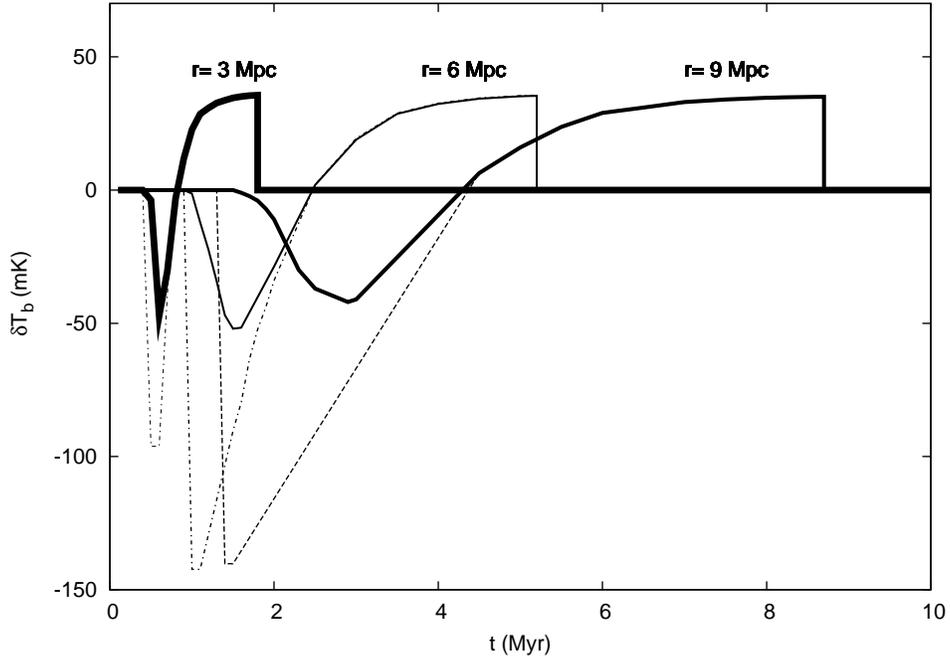}
\caption{The 21 cm brightness temperature $\delta T_b$
as a function of time at three comoving distances $r=3$, 6, 9 Mpc, respectively, for the UV heating model
in Liu et al (2007). The dashed lines are for the solutions
with time-independent W-F coupling, while solid lines show those in
which the time dependence of the W-F coupling is considered.
}
\label{fig12}
\end{center}
\end{figure}

The time evolution of W-F coupling shown in \S 4 should be
considered in calculating the 21 cm emission and absorption from the
halo around the first sources. There are many self-consistent models
with different parameters, such as the UV photon luminosity,
frequency spectrum, the index of power law spectrum, the temperature
of black-body spectrum, etc. (e.g. Cen 2006, Liu et al. 2007, Chen
\& Miralda-Escude 2008). But no one of these models considered the
time-dependence of the W-F coupling.

To show the importance of the time-evolution of the W-F coupling, we
re-calculate the brightness temperature $\delta T_b$ of 21 cm
signals of one of the models developed in Liu et al. (2007) where
source intensity is $\dot{E}=7.25\times 10^{44} erg s^{-1}$ as in
their Figs. 3 and 4, which is self-consistent in terms of heating
and cooling of gas. The results are presented in Figure 12.

Figure 12 shows the time-dependence of the 21 cm brightness
temperature at three shells with radius 3, 6, and 9 h$^{-1}$ Mpc.
For each radius, there are two curves, one does not consider the
time-dependence of the W-F coupling, and one does. It shows that the
brightness temperatures are affected by the time evolution of the
W-F coupling significantly. The time dependence of the W-F coupling
will make the 21 cm signals weaker, especially for absorption
features by a factor 2-5. This is because absorption features are
formed just after the light front passes a particular location, at a
time when the intensity of Ly$\alpha$ photons in the Boltzmann
distribution is far lower than their saturated values (Fig. 6). The
absorption areas of the 21 cm signal would suffer more from
insufficient Ly$\alpha$ photons because they are too close to the
light front. On the other hand, the decrease in $\delta T_b$ is
negligible for emission areas.

\section{Discussion and conclusions}

The kinetics of Ly$\alpha$ resonant photons in the HI media with high
optical depth $\tau$ can basically be described as diffusion in both
the physical space and the frequency space. If Ly$\alpha$ photons do not
join the diffusion in the frequency space, the transfer of
Ly$\alpha$ photons in the physical space is very inefficient, as the
number of scattering needed for escape is proportional to $\tau^2$. The
resonant scattering of Ly$\alpha$ photons and
neutral hydrogen makes the diffusion processes in the physical space
coupled to the diffusion processes in the frequency space. First, the
diffusion in the frequency space provides a shortcut for the diffusion
in the physical space. It makes the mean number of scattering for escape
to be approximately proportional to $\tau$. Second, the
bounce back of resonant scattering provides a mechanism of quickly
restoring $\nu_0$ photons from $x \simeq \pm (2-3)$ photons.
Finally the W-F coupling is realized simultaneously with the
restoration of the $x=0$ photons.

The mechanism of ``escape via shortcut'' plus ``bounce back'' is
mainly carried out by the photons with frequency $x \simeq  \pm (2-3)$.
In a 21 cm emission region of physical size $R\simeq 10$ kpc and $f_{\rm HI}>0.1$,
the optical depth of $x \simeq \pm (2-3)$ photons is still larger
than 1. Therefore, it is reasonable to use Eddington approximation.
On the other hand, the optical depth of $x \simeq \pm (2-3)$ photons
is much less than that of the $x=0$ photons. The $x \simeq  \pm (2-3)$
photons can transfer and enter the 21 cm emission region in a time
scale less than $10^5$ years. Therefore, the mechanism of ``escape
via shortcut'' plus ``bounce back'' is able to timely support the
W-F coupling of the 21 cm emission shell with Ly$\alpha$ photons
from the center objects.

The time dependence of the W-F coupling would make the 21 cm signals
weaker than the predication given by models which do not consider this time-evolution.
Especially at the early stage of the formation of the 21 cm signal
regions, the intensity of the local Boltzmann distribution is still very low,
and therefore, one cannot assume that the spin temperature of 21 cm
is locked to the kinetic temperature of gas. It may yield a low brightness
temperature of the 21 cm signals.

Although the mechanism of ``escape via shortcut'' plus ``bounce back''
helps Ly$\alpha$ diffusion, it does not mean that this mechanism will
reduce the Gunn-Peterson optical depth of the Ly$\alpha$ photons. On the
contrary, the resonant scattering will lead to a slight increase of the
optical depth of the Ly$\alpha$ in the 21 cm region, as the resonant scattering
impedes the cosmic redshift (Figure \ref{fig2}). Consequently, there should be
no observable redshifted optical signal with the frequency $\nu_0/(1+z)$
to be spatially correlated with the $(1+z)$ redshifted 21 cm signal.

The evolution of photons described by eqs.(\ref{eq16}) and
(\ref{eq17}) conserves photon numbers.  The number of
Ly$\alpha$ photons is basically conserved if one can ignore the
Ly$\alpha$ photon destruction processes, such as the two-photon process
(Spitzer \& Greenstein 1951, Osterbrock, 1962). Thus, subsequent
evolution of the Ly$\alpha$ photons in the 21 cm region is to
diffuse to a large sphere around the first stars. At the same time,
the Ly$\alpha$ photons will be redshifted. When the redshift is
large enough, their Gunn-Peterson optical depth will be small, and
finally these photons will escape from the halo (e.g. Miralda-Escude \& Rees 1998;
Loeb \& Rybicki 1999; Zheng \& Miralda-Escude 2002; Haiman \& Cen 2005). The
escaping sphere should be larger than the size of the 21 cm region. Therefore,
redshifted Ly$\alpha$ optical signal with low surface brightness may come from
a big halo around the 21 cm emission region.

\acknowledgments

This work is supported in part by the US NSF under the grants
AST-0506734 and AST-0507340 and by ARO grant W911NF-08-1-0520. WX is
grateful for the hospitality of National Astronomical Observatories
of China, where part of this work was done.

\appendix
\section{Numerical algorithm}

To solve equations (\ref{eq16}) and (\ref{eq17}) as a system, our computational
domain is $(r, x) \in [0,r_{max}]\times[x_{left},x_{right}]$, where $r_{max}$,
$x_{left}$ and $x_{right}$ are chosen such that the solution vanishes to zero
outside the boundaries. In the following, we describe numerical techniques
involved in our algorithm, including approximations to the spatial derivatives,
integrals in the frequency domain, numerical boundary condition and time evolution.

\subsection{The WENO algorithm: approximations to the spatial derivatives}

The spatial derivative terms in equations (\ref{eq16}) and (\ref{eq17}) are approximated
by a fifth order finite difference WENO scheme.

We first give the WENO reconstruction procedure in approximating
$\frac{\partial j}{\partial x}$,
\begin{equation}
\frac{\partial j(\eta^n,r_i,x_j)}{\partial x} \approx \frac{1}{\Delta x}
(\hat{h}_{j+1/2}-\hat{h}_{j-1/2}),
\end{equation}
with fixed $\eta = \eta^n$ and $r = r_i$.
The numerical flux $\hat{h}_{j+1/2}$ is obtained by the fifth order
WENO approximation in an upwind fashion, because the wind direction is fixed (negative).
Denote
\begin{equation}
h_j = j(\eta^n,r_i,x_j),    \hspace{25mm} j = -2, -1,\cdots,N+3\\
\end{equation}
with fixed $n$ and $i$. The numerical flux from the WENO procedure is
obtained by
\begin{equation}
\hat{h}_{j+1/2}=\omega_1\hat{h}_{j+1/2}^{(1)}+\omega_2\hat{h}_{j+1/2}^{(2)} +
\omega_3\hat{h}_{j+1/2}^{(3)},\\
\end{equation}
where $\hat{h}_{j+1/2}^{(m)}$ are the three third order fluxes on
three different stencils given by
\begin{eqnarray*}
\hat{h}_{j+1/2}^{(1)} &=& -\frac{1}{6}h_{j-1}+\frac{5}{6}h_{j}+\frac{1}{3}h_{j+1},\\
\hat{h}_{j+1/2}^{(2)} &=& \frac{1}{3}h_{j}+\frac{5}{6}h_{j+1}-\frac{1}{6}h_{j+2},\\
\hat{h}_{j+1/2}^{(3)} &=&
\frac{11}{6}h_{j+1}-\frac{7}{6}h_{j+2}+\frac{1}{3}h_{j+3},
\end{eqnarray*}
and the nonlinear weights $\omega_m$ are given by,
\begin{equation}
\omega_m =
\frac{\check{\omega}_m}{\displaystyle\sum_{l=1}^3\check{\omega}_l},
\hspace{5mm}
\check{\omega}_l = \frac{\gamma_l}{(\epsilon+\beta_l)^2},\\
\end{equation}
where $\epsilon$ is a parameter to avoid the denominator to become
zero and is taken as $\epsilon = 10^{-8}$. The linear weights
$\gamma_l$ are given by
\begin{equation}
\gamma_{1} = \frac{3}{10},\hspace{3mm} \gamma_{2} = \frac{3}{5},
\hspace{3mm} \gamma_{3} = \frac{1}{10},
\end{equation}
and the smoothness indicators $\beta_{l}$ are given by,
\begin{eqnarray*}
\beta_1 &=& \frac{13}{12}(h_{j-1}-2h_{j}+h_{j+1})^2 +\frac{1}{4}(h_{j-1}-4h_{j}+3h_{j+1})^2,\\
\beta_2 &=& \frac{13}{12}(h_{j}-2h_{j+1}+h_{j+2})^2 +\frac{1}{4}(h_{j}-h_{j+2})^2,\\
\beta_3 &=& \frac{13}{12}(h_{j+1}-2h_{j+2}+h_{j+3})^2
+\frac{1}{4}(3h_{j+1}-4h_{j+2}+h_{j+3})^2.
\end{eqnarray*}

To approximate the $r$-derivatives in the system of equations (\ref{eq16}) and
(\ref{eq17}), we need to perform the WENO procedure based on a characteristic
decomposition. We write
the left hand side of equations (\ref{eq16}) and (\ref{eq17}) as
\begin{equation}
{\bf u}_t + A {\bf u}_r  \\
\end{equation}
where ${\bf u} = (j, f)^T$ and
\[
A =\left( \begin{array}{cc}
0 & 1 \\
\frac{1}{3} & 0\end{array} \right)\]
is a constant matrix.
To perform the characteristic decomposition,
we first compute the eigenvalues, the right eigenvectors, and the left eigenvectors
of $A$ and denote them by,
$\Lambda$, $R$ and $R^{-1}$.
We then project ${\bf u}$ to the local characteristic fields ${\bf v}$ with ${\bf v}
 = R^{-1}{\bf u}$.
Now ${\bf u}_t + A {\bf u}_r$ of the original system is decoupled as two independent
equations as ${\bf v}_t + \Lambda {\bf v}_r$.
We approximate the derivative ${\bf v}_r$ component by component, each with the correct upwind
direction, with the WENO reconstruction procedure
similar to the procedure described above for $\frac{\partial j}{\partial x}$.
In the end, we transform ${\bf v}_r$ back to the physical space by ${\bf u}_r = R {\bf v}_r$.
We refer the readers to Cockburn et al. 1998 for more implementation details.

\subsection{Normalization of the re-distribution term}

We apply the rectangular rule to evaluate $\int R(x, x') j dx$. The rectangular rule is known
to have spectral accuracy for smooth integrated function $R(x, x')j(x')$ with compact
support.
In equations (\ref{eq16}) and (\ref{eq17}), it is known that
\begin{equation}
\int R(x, x')dx' = \phi(x) ,
\end{equation}
which is crucial for photon conservation. However,
\[
\sum_{j} R(x, x_j)  \Delta x = \phi(x)
\]
may not be true in general.
To numerically preserve the photon conservation, we first compute
\begin{equation}
ratio(x, \Delta x) = \frac{\phi(x)}{ \sum_j R(x, x_j) \Delta x},
\end{equation}
then we normalize the collision term by approximating $\int R(x, x')j(t, r, x')dx'$ with
\begin{equation}
ratio(x, \Delta x ) \, \sum_j R(x, x_j) j(\eta,r,x_j) \Delta x
\end{equation}
with fixed $\eta$ and $r$.

\subsection{Implementation of the boundary condition}

 The source term given in the equation (\ref{eq16}) is implemented as a boundary
condition on $f(\eta,r=r_0,x)$.
\[
f(\eta,r=r_0,x) = s_0\phi_s(x)
\]
For the intensity $j$, a reflective boundary condition is used at $r = r_0$. At the
boundary of $r =r_{max}$, $x = x_{left}$ and $x = x_{right}$, we use zero boundary
conditions for both $j$ and $f$, because of the way we choose $r_{max}$, $x_{left}$
and $x_{right}$.

\subsection{Time evolution}

The time derivatives  $\frac{\partial j}{\partial \eta}$  and
$\frac{\partial f}{\partial \eta}$ are approximated by the third-order TVD Runge Kutta time
discretization (Shu \& Osher, 1988). For systems of ODEs $u_t =
L(u)$, the third order Runge-Kutta method is
\begin{eqnarray*}
u^{(1)} &=& u^n + \Delta t L(u^n,t^n),\\
u^{(2)} &=& \frac{3}{4}u^n + \frac{1}{4}(u^{(1)}+\Delta t
L(u^{(1)},t^n +
\Delta t)),\\
u^{n+1} &=& \frac{1}{3}u^n + \frac{2}{3}(u^{(2)}+\Delta t
L(u^{(2)},t^n + \frac12 \Delta t)).
\end{eqnarray*}

\subsection{Test with the conservation of the photon number}

 From eq.(\ref{eq16}) we have
\begin{equation}
\frac{\partial}{\partial \eta} \int j dx + \frac{\partial }
{\partial r} \int f dx = 0
\end{equation}
Therefore, for time-independent solution we have
$\int f(r,x)dx = {\rm const}$. It yields
\begin{equation}
\int f(r,x)dx = \int f(0,x)dx= S_0
\end{equation}
That is, the flux at saturated states is $r$-independent. This is the conservation of
the number of photons. It can be used to test the algorithm.

\end{document}